\documentclass[aip,preprint]{revtex4-1}

\usepackage{lineno,hyperref}
\usepackage{amsmath}
\usepackage{amssymb}
\usepackage{threeparttable}
\usepackage{multirow}
\usepackage{array}
\usepackage{color}
\usepackage{graphicx}
\usepackage{subfigure}
\usepackage{geometry}
\usepackage{float}
\usepackage{framed}
\usepackage{booktabs}
\usepackage{bm}
\usepackage{dcolumn}

\renewcommand{\vec}[1]{\bm{#1}}

\begin{document}
	

\title{Unified Gas-kinetic Scheme with Multigrid Convergence for Rarefied Flow Study}

\author{Yajun Zhu}
\email{zhuyajun@mail.nwpu.edu.cn}
\affiliation{National Key Laboratory of Science and Technology on Aerodynamic Design and Research, Northwestern Polytechnical University, Xi'an, Shaanxi 710072, China}
\author{Chengwen Zhong}
\email{zhongcw@nwpu.edu.cn}
\affiliation{National Key Laboratory of Science and Technology on Aerodynamic Design and Research, Northwestern Polytechnical University, Xi'an, Shaanxi 710072, China}
\author{Kun Xu}
\email{makxu@ust.hk}
\affiliation{Department of Mathematics, Hong Kong University of Science and Technology, Hong Kong, China}

\date{\today}

\begin{abstract}
The unified gas kinetic scheme (UGKS) is a direct modeling method based on the gas dynamical model on the mesh size and time step scales. With the implementation of particle transport and collision in a time-dependent flux function, the UGKS can recover multiple flow physics from the kinetic particle transport to the hydrodynamic wave propagation. In comparison with direct simulation Monte Carlo (DSMC), the equations-based UGKS can use the implicit techniques in the updates of macroscopic conservative variables and microscopic distribution function. The implicit UGKS significantly increases the convergence speed for steady flow computations, especially in the highly rarefied and near continuum regime. In order to further improve the computational efficiency, for the first time a geometric multigrid technique is introduced into the implicit UGKS, where the prediction step for the equilibrium state and the evolution step for the distribution function are both treated with multigrid acceleration. More specifically, a full approximate nonlinear system is employed in the prediction step for fast evaluation of the equilibrium state, and a correction linear equation is used in the evolution step for the update of the gas distribution function. As a result, convergent speed has been greatly improved in all flow regimes from rarefied to the continuum ones. The multigrid implicit UGKS (MIUGKS) is used in the non-equilibrium flow study, which includes microflow, such as lid-driven cavity flow and the flow passing through a finite-length flat plate, and high speed one, such as supersonic flow over a square cylinder. The MIUGKS shows $5$ to $9$ times efficiency increase over the previous implicit scheme. For the low speed microflow, the efficiency of MIUGKS is several orders of magnitude higher than the DSMC. Even for the hypersonic flow at Mach number $5$ and Knudsen number $0.1$, the MIUGKS is still more than $100$ times faster than the DSMC method for a convergent steady state solution.
\end{abstract}


\maketitle
\section{Introduction}\label{sec:introduction}
With discretized particle velocity space, the unified gas kinetic scheme (UGKS) was an extension from the gas kinetic scheme (GKS) for the Navier-Stokes (NS) solution to the flow dynamics in the entire Knudsen regimes \cite{xu2001gas,xu2010unified}. As a NS solver, the GKS updates macroscopic conservative flow variables only. But, the UGKS evolves both macroscopic flow variables and the microscopic gas distribution function. In both schemes, a time-dependent gas distribution function from kinetic model equation is used for the flux evaluation, and this time evolving solution covers the gas dynamics from the initial non-equilibrium state to the final hydrodynamic equilibrium one. The real solution used for the updates of macroscopic flow variables and the gas distribution function depends on the relative values of particle collision time $\tau$ and the local numerical time step $\Delta t$. The main difference between GKS and UGKS is the set-up of the initial distribution function around a cell interface at the beginning of each time step. For GKS, it is reconstructed from the updated macroscopic flow variables through the Chapman-Enskog expansion, and for UGKS it directly uses the updated gas distribution function. The UGKS can describe highly non-equilibrium flow physics due to the update of the discretized distribution function. With a variation of the ratio between the time step and local particle mean collision time, the UGKS is capable to present the Boltzmann solution in the rarefied flow regime and the NS solution in the continuum flow domain. In the transition regime, a reliable solution can be obtained by UGKS as well \cite{xu-liu}. In addition, different from the discrete velocity method (DVM) \cite{mieussens2000discrete} and the direct simulation of Monte Carlo (DSMC) method \cite{bird1994book}, the cell size and time step in UGKS are not restricted by the particle mean free path and collision time due to the implicit treatment of particle collision term inside each cell with the help of updated macroscopic flow variables. The distinguishable multi-scale feature of the UGKS makes it suitable in the gas dynamics study with multiple flow regimes in a single computation, such as the flow passing through a nozzle \cite{chen2012adaptive} from the inside highly compressed continuum flow ($\tau/\Delta t \ll 1$) to the outside highly rarefied one ($\tau/\Delta t \gg 1$).

In the past years, the UGKS has been validated extensively and it gives accurate solutions in all flow regimes \cite{xu2015book}. The method can be easily extended to more complex gases, such as diatomic molecule gas \cite{liu2014diatomic} and multi-species flow \cite{wang2014multispecies}. The methodology of direct modeling in UGKS can be used to construct numerical methods in other transport processes, such as radiation and phonon transfer \cite{sun2015radiative,guo2016phonon} and plasma physics \cite{liu2016plasma}. The UGKS provides a promising tool and shows great potentials in the engineering applications, e.g., to the micro-electro-mechanical system and spacecraft designs.

The barriers in front of UGKS for preventing its wide applications in comparison with the DSMC method are the high memory requirements and computational cost, especially for the hypersonic flow and high temperature variation. However, since the UGKS is still an equation-based method, it has advantages in comparison with particle methods in the reduction of computational cost. Many acceleration techniques in traditional computational fluid dynamics (CFD) can be directly adopted in UGKS. One way to reduce the computational cost in UGKS is to reduce the discretization points, such as adopting a moving and adaptive mesh in the physical and velocity spaces \cite{chen2012adaptive}. With adaptive discretization techniques, the computational cost could be controlled to a tolerable level even for highly non-equilibrium flow problems. Another way is to adopt acceleration techniques. For explicit scheme, the numerical stability imposes the Courant-Friedrichs-Lewy (CFL) condition on the time step. But with implicit treatments, this constraint can be released and the computational efficiency can be greatly enhanced. The implicit GKS has been constructed for a faster convergence to the Navier-Stokes solutions \cite{li2006applications,xu2005multidimensional,jiang2012implicit,li2014implicit}. For rarefied flows, several implicit schemes have been proposed based on the iterative algorithms in updating the discretized gas distribution function \cite{yang1995rarefied,mao2015study}. As pointed out in  \cite{mieussens2000implicit, mieussens2000discrete}, the direct explicit treatment of the equilibrium state in the collision term of the kinetic equation may slow down the convergence of the implicit schemes, especially near continuum flow regime. Hence in a previous study \cite{zhu2016implicit}, an implicit UGKS with a prediction step for the equilibrium state was developed to increase acceleration convergence. By first updating the conservative variables implicitly, the collision term in the kinetic equation can be treated in an implicit way, which drives the gas distribution function to a steady state solution efficiently. The implicit UGKS has been validated to be a robust and efficient method in all flow regimes. In order to further speed up the convergence of UGKS for a steady flow solution, the multigrid method, which is one of the most outstanding acceleration techniques in CFD, will be implemented in the implicit UGKS in this paper.

The study of multigrid technique may originate from 1960s \cite{fedorenko1962relaxation,fedorenko1964speed}. Since Brandt's works \cite{brandt1977multi} in 1970s, the multigrid method got a fast development in practical computations. Now the multigrid method is commonly used in CFD community \cite{blazek2001book} for solving the Euler and Navier-Stokes equations \cite{jameson1983solution,yoon1987lu}. It has been applied to the GKS \cite{xu1994,jiang2012implicit} for acceleration to the steady state solutions for the continuum flow computations. There are many monographs \cite{brandt2011guide,trottenberg2000multigrid,stuben1982multigrid,wesseling1991introduction} about multigrid techniques and their numerical implementations. The basic idea behind all multigrid strategies is to accelerate the solution at fine grid by computing corrections on a coarser grid \cite{mavriplis1995multigrid} to eliminate low-frequency errors efficiently. In general, an iterative algorithm can reduce the high-frequency errors faster than the low-frequency ones. The multiple grid method is to make the transition between the low and high frequency modes through a change of cell size, and to eliminate the low frequency error in an even coarse mesh by increasing its spatial frequency.

In this paper, we develop a multigrid method for the implicit UGKS, which further improves its convergence efficiency in rarefied flow computations. The implicit UGKS \cite{zhu2016implicit} has the prediction stage for evaluating the implicit part of equilibrium state in the collision term, and the evolution stage for updating the gas distribution function. Both stages of the implicit UGKS are treated with multigrid techniques to ensure a fast convergence in all flow regimes. It turns out that the macroscopic equations become a nonlinear system, while the implicit evolution equations for the distribution function at discrete particle velocities are still linear ones. As a result, the full approximation storage scheme (FAS) \cite{brandt2011guide} is used in the prediction step for the conservative flow variables and the correction scheme (CS) \cite{brandt2011guide,trottenberg2000multigrid} for solving linear equations is utilized in the evolution of the distribution function. For the first time, a multigrid method is used in the UGKS for the rarefied flow computation. After presenting the scheme, many rarefied flow cases from low to high speed ones covering a wide range of flow regimes will be studied, such as lid-driven cavity flow, flow passing through a finite-length flat plate, and supersonic flow over a square cylinder. In all cases presented in the current paper, the implicit UGKS with multigrid acceleration is much more efficient than the DSMC method with orders of magnitude differences.

The paper is organized as follows. In section \ref{sec:numericalMethod}, the multigrid implicit UGKS (MIUGKS) is presented. Section \ref{sec:remarks} is about the analysis and remarks on the current multigrid method. Section \ref{sec:numericalTests} presents the rarefied flow studies using the MIUGKS. The last section is the conclusion.

\section{Multigrid implicit UGKS}\label{sec:numericalMethod}
In this section, the implicit UGKS will be introduced first \cite{zhu2016implicit}. The implicit UGKS is a pseudo-time-marching scheme for steady state solution. In fact, the implicit scheme can be interpreted as a numerical smoothing method, which can be naturally incorporated into a multigrid framework. The basic components in the multigrid method will be described via the detailed formulation of a two-grid cycle. The extension to multiple grids is straight forward through a recursive way on the basis of the two-grid cycle.

\subsection{Implicit UGKS}\label{sec:IUGKS}
For steady flows, the governing equation of macroscopic variables averaged in a finite volume $i$ gives
\begin{equation}\label{eq:macroSteadyGov}
\dfrac{1}{V_i} \sum_{j \in N(i)}{S_{ij}  {\vec{F}}_{ij}} = \vec{0},
\end{equation}
where $N(i)$ is the set of neighbors of cell $i$, and $j$ is one of the neighboring cell, and $ij$ denotes the interface between cells $i$ and $j$. Here $S_{ij}$ is the area of the interface $ij$ and $V_i$ is the volume of the cell $i$. $\vec{F}_{ij}$ are the fluxes of conservative variables $\vec{W} = \left(\rho, \rho U, \rho V, \rho \varepsilon \right)^T $ passing through the cell interface $ij$. Eq.~(\ref{eq:macroSteadyGov}) describes the balance of interface fluxes for cell $i$ at steady state.

For the gas distribution function $f_{i,k}$ at the discretized velocity $\vec{u}_k$, the governing equation can be written as
\begin{equation}\label{eq:microSteadyGov}
\dfrac{1}{V_i}\sum_{j\in N(i)}{S_{ij} u_{k,n} {\tilde f}_{ij,k}} - \dfrac{{g}_{i,k}-{f}_{i,k}}{\tau_i} = 0,
\end{equation}
where $u_{k,n}$ is the normal component of $\vec{u}_k$ along the interface $ij$. The interface distribution function ${\tilde f}_{ij,k}$ is a local physical time $\Delta t_p$ averaged quantity, which identifies different physics in different regimes. The equilibrium state $g_{i,k}$ can be the Maxwellian distribution, or a Shakhov-type model with the justification of the Prandtl number. The multiple scale nature of UGKS is fully determined by the modeling of the flux function at a cell interface ${\tilde f}_{ij,k}$, which will be presented later.

Since Eq.~(\ref{eq:macroSteadyGov}) and Eq.~(\ref{eq:microSteadyGov}) depict the final steady state solution, which denotes time $t \to \infty$ for explicit scheme or iteration step $n \to \infty$ for iterative methods, they could be directly regarded as the implicit governing equations of the accurate final solution. In general, the solution is basically impossible to be obtained in one step. Numerical computation will start from an approximate solution (or initial state solution) and then get more accurate solutions step by step using explicit time-marching schemes or implicit iterative methods.

For the implicit UGKS, given an approximate solution $f_{i,k}^n$ and $\vec{W}_i^{n}$ at step $n$, the errors for macroscopic and microscopic variables can be defined as
\begin{equation}\label{eq:macroErrorDef}
{\vec E}_{i}^{n} = {\vec W}_i -{\vec W}_{i}^{n},
\end{equation}
and
\begin{equation}\label{eq:microErrorDef}
e_{i,k}^n = f_{i,k} - f_{i,k}^n.
\end{equation}
The residuals become
\begin{equation}\label{eq:macroResidualDef}
{\vec R}_i^n = -\dfrac{1}{V_i}\sum_{j\in N(i)}{S_{ij}{\vec F}_{ij}^{n}},
\end{equation}
and
\begin{equation}\label{eq:microResidualDef}
r_{i,k}^n = \dfrac{{g}_{i,k}-{f}_{i,k}^n}{\tau_i} - \dfrac{1}{V_i}\sum_{j\in N(i)}{S_{ij} u_{k,n} {\tilde f}_{ij,k}^n}.
\end{equation}
As a result, the residual equations (or defect equations) for implicit iterations go to
\begin{equation}\label{eq:macroDefectDef}
\dfrac{1}{V_i} \sum_{j \in N(i)}{ S_{ij} \left( \vec{F}_{ij}-\vec{F}_{ij}^n \right)} = \vec{R}_i^n,
\end{equation}
and
\begin{equation}\label{eq:microDefectDef}
\dfrac{e_{i,k}^n}{\tau_i} +\dfrac{1}{V_i}\sum_{j\in N(i)}{S_{ij} u_{k,n} {e}_{ij,k}^n}= r_{i,k}^n.
\end{equation}

If ${\vec E}_{i}^{n}$ and $e_{i,k}^{n}$ were precisely solved from Eq.~(\ref{eq:macroDefectDef}) and Eq.~(\ref{eq:microDefectDef}), we could get the exact solution ${\vec W}_{i}$ and $f_{i,k}$ from Eq.~(\ref{eq:macroErrorDef}) and Eq.~(\ref{eq:microErrorDef}). However, it is much too difficult to solve the residual equations (\ref{eq:macroDefectDef}) and (\ref{eq:microDefectDef}) with the full UGKS terms of ${\vec{F}_{ij}-{\vec{F}_{ij}^n}}$ and $e_{ij,k}^n$. Moreover, it requires the unknown equilibrium state $g_{i,k}$ in evaluation of microscopic residual $r_{i,k}^n$. Therefore, we divide the solving process into two steps, i.e., prediction for equilibrium state and evolution of gas distribution function. In the prediction step, with the simplified implicit fluxes on the left hand side of Eq.~(\ref{eq:macroDefectDef}) we can approximately solve the Eq.~(\ref{eq:macroDefectDef}) to get a correction of conservative variables $\Delta \vec{W}_{i}^n$ as an approximation of ${\vec E}_{i}^{n}$. Then the equilibrium state $g_{i,k}$ in the evaluation of $r_{i,k}^n$ can be approximated by $\tilde{g}_{i,k}^{n+1}$ obtained from $\tilde{\vec{W}}_{i}^{n+1} = \vec{W}_{i}^n +\Delta \vec{W}_{i}^n$. Consequently, in the evolution step we can obtain a correction of distribution function $\Delta f_{i,k}^{n}$ as an approximate $e_{i,k}^n$ once we solve Eq.~(\ref{eq:microDefectDef}) using simplified fluxes ${e}_{ij,k}^n$. Then the distribution function can be updated by $f_{i,k}^{n+1} = f_{i,k}^n +\Delta f_{i,k}^n$ and the equilibrium state $g_{i,k}^{n+1}$ and the conservative variables $\vec{W}_{i}^{n+1}$ can be renewed by the compatibility condition from $f_{i,k}^{n+1}$. Following these procedures iteratively, the convergent solution can be obtained step by step from the corrections, accompanied with error smoothing and reduction. Details in these two steps will be introduced in the following.

\subsubsection{Prediction step for equilibrium state }\label{sec:prediction}
In order to evaluate the residuals $r_{i,k}^n$ in Eq.~(\ref{eq:microResidualDef}), the equilibrium state $g_{i,k}$ should be given first. Here we give a predicted one $\tilde{g}_{i,k}$ by solving the implicit governing equations (\ref{eq:macroDefectDef}) of macroscopic variables.

For evaluation of the residual $ {\vec R}_i^n $ in Eq.~(\ref{eq:macroResidualDef}), we use
\begin{equation}\label{eq:macroResidualCalc}
{\vec F}_{ij}^{n}
={\sum_k{ \dfrac{1}{\Delta t_p} \int_0^{\Delta t_p} { u_{k,n} {\tilde f}_{ij,k}^n(t) {\vec \psi_k} dt}}},
\end{equation}
where ${\tilde f}_{ij,k}^n(t)$ is a time-dependent distribution function constructed by the analytic solution of the kinetic model equation along a characteristic line, and $\vec \psi_k $ is the vector for its moments of mass, momentum and energy. $\Delta t_p$ is the physical time step determined by the CFL condition with a Courant number less than $1$, which recovers the local flow physics. The residuals $\vec{R}_i^n$ are completely evaluated by explicit UGKS fluxes, see details in papers \cite{xu2010unified,huang2012unified,xu2015book}. In order to solve Eq.~(\ref{eq:macroDefectDef}),the fluxes on the left hand side will be simplified by Euler equations-based flux splitting method,
\begin{equation}\label{eq:macroFluxSplitting}
\begin{aligned}
{\vec{F}_{ij}-{\vec F}_{ij}^{n}}  = & \dfrac{1}{2} \left[ {\vec T}_i+ {\vec T}_j 
+  \Gamma_{ij}\left({{\vec W}_{i}}-{ {\vec W}_{j}}\right)\right]\\
& - \dfrac{1}{2} \left[{ {\vec T}_{i}^{n}}+{ {\vec T}_{j}^{n}} 
+  \Gamma_{ij}\left({{\vec W}_{i}^{n}}-{{\vec W}_{j}^{n}}\right)\right],
\end{aligned}
\end{equation}
where $\vec{T}$ is the Euler flux.
$\Gamma_{ij}$ satisfies
\begin{displaymath}
 \Gamma_{ij} \ge  \Lambda_{ij} = \left| {{\vec U}_{ij}  \cdot {\vec n}_{ij} } \right| + a_s,
\end{displaymath}
where $\Lambda_{ij}$ represents the spectral radius of the Euler flux Jacobian, which can be evaluated by the macroscopic velocity $\vec{U}_{ij}$ and speed of sound $a_s$ at the interface $ij$. Here $\vec{n}_{ij}$ is the normal vector of the interface along the direction from cell $i$ to cell $j$. Generally, a stable factor $s_{ij}$ \cite{li2006applications,chen2000fast} related to the kinematic viscosity coefficient $\nu$ can be introduced into the calculation of $\Gamma_{ij}$,
\begin{equation}\label{eq:predictionSpectralRadius}
\Gamma_{ij} = \Lambda_{ij} + s_{ij} = \Lambda_{ij} + \dfrac{2 \nu}{\vec{n}_{ij} \cdot \left(\vec{x}_j - \vec{x}_i\right) }.
\end{equation}
Then the residual equations can be rewritten as
\begin{widetext}
\begin{equation}\label{eq:macroFinalDefect}
\dfrac{1}{2 V_i} \sum_{j\in N(i)}{ S_{ij} \Gamma_{ij} } \vec{E}_i^n
+ \dfrac{1}{2 V_i} \sum_{j\in N(i)}
{ S_{ij} \left[ {\vec T} ({ {\vec W}_{j}^n+\vec{E}_j^n})-{\vec T}({\vec{W}}_{j}^{n}) - \Gamma_{ij} {{\vec E}_{j}^{n}} \right] }
= \vec{R}_i^n.
\end{equation}
\end{widetext}
For two dimensional cases on structured mesh, it will form a penta-diagonal matrix, which can be solved by LU-SGS iterations \cite{jameson1987lower,yoon1988lower,zhu2016implicit}. With the correction $\Delta \vec{W}_i^n$ for conservative variables as an approximation of $ \vec{E}_i^n$, we can get the predicted equilibrium state $\tilde g_{i,k}$ from the newly evolved macroscopic variables $\vec{W}_{i}^{n}+\Delta \vec{W}_i^n$.

\subsubsection{Evolution step for updating particle distribution function}\label{sec:evolution}
Once we get the predicted equilibrium state $\tilde{g}_{i,k}$, the microscopic residual in Eq.~(\ref{eq:microResidualDef}) can be evaluated. Simplifying the numerical flux on the left hand side of Eq.~(\ref{eq:microDefectDef}) by an upwind approach,
\begin{equation}\label{eq:microUpwind}
\begin{aligned}
{e}_{ij,k}^{n}
 = & \dfrac{1}{2} \left( {e}_{i,k}^{n}+ {e}_{j,k}^{n} \right) 
   + \dfrac{1}{2} {\rm sign}({\vec u}_k \cdot {\vec n}_{ij}) \left( {e}_{i,k}^{n}-{e}_{j,k}^{n} \right)\\
= &\dfrac{1}{2}\left[ 1 + {\rm sign}(u_{k,n})\right] {e}_{i,k}^{n} 
+ \dfrac{1}{2}\left[ 1 - {\rm sign}(u_{k,n})\right] {e}_{j,k}^{n},\\
\end{aligned}
\end{equation}
we get the residual equation
\begin{equation}\label{eq:microFinalDefect}
D_{i,k} {e}_{i,k}^{n}
+\sum_{j\in N(i)}{D_{j,k} {e}_{j,k}^{n}}
=r_{i,k}^n,
\end{equation}
where
\begin{equation}\label{eq:evolutionCoefficients}
\begin{aligned}
&D_{i,k} = \dfrac{1}{{\tilde \tau}_i}+\dfrac{1}{2 V_i} \sum_{j\in N(i)}{u_{k,n} S_{ij}  \left[ 1 + {\rm sign}(u_{k,n})\right]},\\
&D_{j,k} = \dfrac{1}{2 V_i} u_{k,n} S_{ij} \left[ 1 - {\rm sign}(u_{k,n})\right].
\end{aligned}
\end{equation}
Here $r_{i,k}^n$ is evaluated by the time-averaged UGKS flux over a physical time step $\Delta t_p$ and the collision term with the predicted equilibrium state $\tilde{g}_{i,k}$. By using LU-SGS iterations, a correction $\Delta f_{i,k}^n$ approximating the error $e_{i,k}^n$ can be obtained from Eq.~(\ref{eq:microFinalDefect}). After obtaining $\Delta f_{i,k}^n$, the solution $f_{i,k}^{n+1}$ can be updated. Consequently, the macroscopic variables can be updated as well by taking moments of the renewed distribution function.

Different from the nonlinear Eq.~(\ref{eq:macroFinalDefect}), Eq.~(\ref{eq:microFinalDefect}) can be regarded as a linear equation for the distribution function error if the mean collision time $\tilde{\tau_i} $ is frozen locally within each iteration step, because the coefficients $D_{i,k}$ and $D_{j,k}$ are only related to the discretization of the physical and velocity space. Therefore, different multigrid techniques are imposed on solving the nonlinear equation (\ref{eq:macroFinalDefect}) of the conservative variables and the linear equation (\ref{eq:microFinalDefect}) of the gas distribution function. Details will be introduced next.

\subsection{A two-grid cycle implicit UGKS}\label{sec:twoGrid}
A two-grid cycle method is a basis for any multigrid algorithm. It is a combination of error smoothing and coarse grid correction. Usually it consists of a pre-smoothing, a coarse grid correction, and a post-smoothing part. Here we implement two-grid cycle technique into the implicit UGKS to develop a multigrid method.

\begin{figure*}
	\centering
	\includegraphics[width=0.96\textwidth]{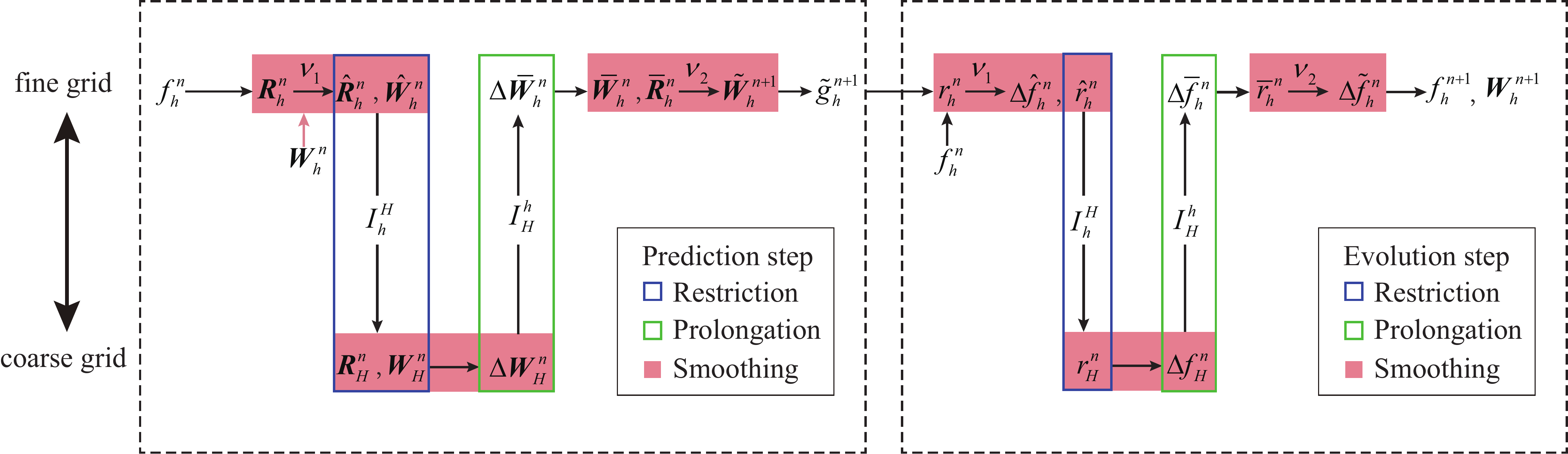}
	\caption{\label{fig:twoCycle}Strucuture of a two-grid cycle for the implicit UGKS with a prediction step (left) and an evolution step (right).}
\end{figure*}

Based on a finer grid $\Omega_h$ and a coarser grid $\Omega_H$, the iteration step of the two-grid cycle for the implicit UGKS is illustrated in Fig.~\ref{fig:twoCycle}, 
with a prediction step and an evolution step. Since interpolations of macroscopic variables and distribution function between two grids are involved, for a better representation the solutions and interpolations will be denoted as grid functions and grid operators respectively. As shown in Fig.~\ref{fig:twoCycle}, the two stages in the implicit UGKS are considered successively with the multigrid technique.

In the prediction step, the residuals $\vec{R}_h^n$ of the implicit macroscopic equations are evaluated by the UGKS fluxes on a fine grid $\Omega_h$ from a given approximate solutions of  $f_h^n$ and $\vec{W}_h^n$. After $\nu_1$ times of pre-smoothing on this level, the residuals and conservative variables are updated to $\hat{\vec{R}}_h^n$ and $\hat{\vec{W}}_h^n$. Then, both the renewed residuals and the smoothed conservative variables are restricted from the fine grid $\Omega_h$ to the coarse grid $\Omega_H$ by a transfer operator $I_h^H$. On a coarse grid, the residual equations with restricted residuals $\vec{R}_H^n$ and $\vec{W}_H^n$ should be solved to get a correction of the conservative variables $\Delta \vec{W}_H^n$. Consequently, the residuals and solutions on the fine grid can be renewed through a prolongated correction $\Delta \bar{\vec{W}}_h^n$ with a transformation operator $I_H^h$. Again, taking $\nu_2$ times for post-smoothing, the smoothed solution ${\tilde{\vec{W}}}_h^{n+1} $ is regarded as the final result in this prediction step to give a predicted equilibrium state for the following evolution of the gas distribution function.

In the evolution step, the residual $r_h^n$ is obtained first on a fine grid $\Omega_h$ from the given approximate solution $f_h^n$ and the predicted equilibrium state $\tilde{g}_h^{n+1}$. Meanwhile, the residual will be renewed after $\nu_1$ times of pre-smoothing, and a correction $\Delta \hat{f}_h^n$ will be obtained during these smoothing processes. As mentioned in Section \ref{sec:evolution}, the residual equation (\ref{eq:microFinalDefect}) of gas distribution function is a linear equation, therefore only the residual $\hat{r}_h^n$ is needed to be restricted on a coarse grid to get a correction. The correction $\Delta f_H^n$ obtained by solving the residual equation on a coarse grid will be prolongated back onto the fine grid. With the interpolated correction $\Delta \bar{f}_h^n$, and renewed residual $\bar{r}_h^n$,  $\nu_2$ times of post-smoothing can be carried out to give an updated distribution function $f_h^{n+1}$. Then the conservative variables $\vec{W}_h^{n+1}$ and equilibrium state $g_h^{n+1}$ can be updated from the moments of the gas distribution function.

So far, the two-grid cycle for the prediction and the evolution steps has been illustrated for the updating of the distribution function from $f_h^n$ to $f_h^{n+1}$. It should be noted that the only difference between the two steps is whether the intermediate smoothed solution is required and restricted on a coarse grid, i.e., the difference between the so-called full approximation storage scheme and correction scheme. In the following, each component of the multigrid method will be introduced in details.

\subsection{Numerical procedures in the multigrid method}\label{sec:component}

\subsubsection{Transfer operator : Restriction}\label{sec:restriction}
For initialization on a successive coarser grid, variables such as the residuals should be transferred (restricted) from finer grid to coarser ones.

\begin{figure}
	\centering
	\includegraphics[width=0.25\textwidth]{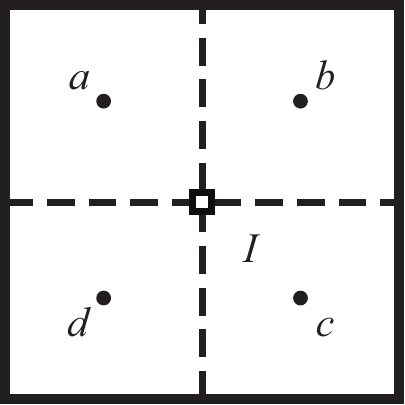}
	\caption{\label{fig:restriction}Stencil for restrictions from fine grids to coarse grids.}
\end{figure}

A restriction operator $I_h^H$ maps fine-grid functions to coarse-grid functions by a volume weighted interpolation for cell-centered schemes. For a specific variable denoted by $Q$, the restricted result inside cell $I$ on coarse grid $\Omega_H$ becomes
\begin{equation}
(I_h^H Q_h)_I = \dfrac{\Sigma_{j \in S(I)} {(Q_h V_h)_j}}{\Sigma_{j\in S(I)} {(V_h)_j}},
\end{equation}
where $S(I)$ is the set of subcells of cell $I$ and $j$ is one of the subcell members. In prediction step, both the smoothed conservative variables $\hat{\vec{W}}_h^n$ and renewed residuals $\hat{\vec{R}}_h^n$ should be restricted to a coarse grid to form the residual equation (\ref{eq:macroFinalDefect}) on $\Omega_H$. 
As illustrated in Fig.~\ref{fig:restriction}, 
we have
\begin{equation}\label{eq:macroRestriction}
\begin{aligned}
(\vec{W}_H^n)_I=(I_h^H \hat{\vec{W}}_h^n)_I &=  \dfrac{1}{V_I} \sum_{j \in S(I)}{(\hat{\vec{W}}_h^n)_j V_j},\\
(\vec{R}_H^n)_I = (I_h^H \hat{\vec{R}}_h^n)_I &= \dfrac{1}{V_I} \sum_{j \in S(I)}{(\hat{\vec{R}}_h^n)_j V_j }
\end{aligned}
\end{equation}
where $S(I)=\{a,b,c,d\}$.
In evolution step, only the residual is needed on a coarse grid, see Eq.~(\ref{eq:microFinalDefect}). Therefore, we have
\begin{equation}\label{eq:microRestriction}
(r_H^n)_I=(I_h^H \hat{r}_h^n)_I = \dfrac{1}{V_I} \sum_{j \in S(I)} {(\hat{r}_h^n)_j V_j}.
\end{equation}

\subsubsection{Smoothing}\label{sec:smoothing}
The smoothing process, whether the pre-smoothing or post-smoothing, is to solve the residual equations
by LU-SGS iterations to get a more accurate solution.
It is a correction process of the solutions on a single grid.
This is why one iteration step of the original implicit UGKS on a single grid is claimed as a smoothing method in Section \ref{sec:IUGKS}.
In the current method, solving the residual equations on a coarsest grid is indeed implemented by applying LU-SGS iterations,
i.e., through adequate times of smoothing.

In the prediction step, the residual equations (\ref{eq:macroFinalDefect}) are rewritten on a coarse grid $\Omega_H$ as
\begin{equation}\label{eq:rewrittenMacroDefect}
\dfrac{1}{V_I} \sum_{J \in N(I)}{S_{IJ}  \vec{F}_{IJ}(\vec{W}_H) }
-\dfrac{1}{V_I} \sum_{J \in N(I)}{S_{IJ}  \vec{F}_{IJ}(\vec{W}_H^{n}) }= ( \vec{R}_H^n)_I.
\end{equation}
where $\vec{W}_H^{n}$ are the restricted conservative variables and $\vec{W}_H $ are the accurate solutions of these equations.
After the first time of LU-SGS iteration on this grid, the above equations can be solved to give new approximation solutions $\vec{W}_H^{(1)}$.
Denoting the $m$-th intermediate solution as $\vec{W}_H^{(m)}$, we get the governing equations for the $(m+1)$-th time of smoothing, i.e.,
\begin{equation}\label{eq:smoothingMacro}
\dfrac{1}{V_I} \sum_{J \in N(I)}{S_{IJ}  \vec{F}_{IJ}}
-\dfrac{1}{V_I} \sum_{J \in N(I)}{S_{IJ}  \vec{F}_{IJ}(\vec{W}_H^{(m)}) }
= (\vec{P}_H^n )_I
 -\dfrac{1}{V_I} \sum_{J \in N(I)}{S_{IJ}  \vec{F}_{IJ}(\vec{W}_H^{(m)}) },
\end{equation}
where $\vec{P}_H^n$ is the forcing function defined as the difference between the residuals directly transferred from the fine grid,
and the macroscopic evolution equations-determined residuals which are recomputed on a coarse grid, i.e.,
\begin{equation}\label{eq:forcingFunction}
(\vec{P}_H^n)_I = (\vec{R}_H^n)_I - \vec{R}_I(\vec{W}_H^n)= (I_h^H \vec{R}_h^n )_I+ \dfrac{1}{V_I} \sum_{J \in N(I)}{S_{IJ} \vec{F}_{IJ}(\vec{W}_H^{n})}.
\end{equation}
This is commonly used in solving nonlinear residual equations \cite{jameson1987lower}.
Here $\vec{F}_{IJ}(\vec{W}_H^{n})$ are calculated by a flux splitting method as in Eq.~(\ref{eq:macroFluxSplitting}).
Therefore, for the $(m+1)$-th smoothing process, the residuals on the right hand side of Eq.~(\ref{eq:smoothingMacro}) can be updated by
\begin{equation}\label{eq:renewResiduals}
\vec{R}_H^{(m)} =   \vec{P}_H+\vec{R}(\vec{W}_H^{(m)}).
\end{equation}
For the first smoothing iteration the Eq.~(\ref{eq:smoothingMacro}) is identical to the Eq.~(\ref{eq:rewrittenMacroDefect}).
By solving Eq.~(\ref{eq:smoothingMacro}) with LU-SGS iterations, multiple smoothing processes can be carried out.

Similarly, for the $(m+1)$-th smoothing process of the distribution function in the evolution step,
the residual equation (\ref{eq:microFinalDefect}) can be rewritten on a coarse grid as
\begin{equation}\label{eq:rewrittenMicroDefect}
D_{I} {e}_{I}^{(m)}
+\sum_{J\in N(I)}{D_{J} {e}_{J}^{(m)}}
=(I_h^H r_{h}^n)_I-\left(D_{I} \Delta{f}_{I}^{(m)}
+\sum_{J\in N(I)}{D_{J} \Delta{f}_{J}^{(m)}}\right),
\end{equation}
where $e_I^{(m)} = f_I- f_I^{(m)}$ and $\Delta{f}_I^{(m)} = f_I^{(m)}-f_I^{(0)}$. Here $f_I$ is the accurate solution of Eq.~(\ref{eq:rewrittenMicroDefect}) and $f_I^{(0)}$ is the initial distribution function on $\Omega_H$
imaginarily restricted from $\Omega_h$.
The purpose for us to give the expression of $e_I^{(m)}$ and $\Delta{f}_I^{(m)}$ by the intermediate distribution function $f_I^{(m)}$
is just for a better understanding of Eq.~(\ref{eq:rewrittenMicroDefect}).
It should be noted that the distribution function $f_I$ is indeed not a necessity in computations,
while only  $\Delta$-quantities are actually involved.
In computation, the residual on the right hand side of Eq.~(\ref{eq:rewrittenMicroDefect}) for each smoothing process is updated by
\begin{equation}\label{eq:renewResidualMicro}
(r_H^{(m)})_I = (I_h^H r_{h}^n)_I-\sum_{\alpha=1}^{m}{\left(D_{I} \delta{f}_{I}^{(\alpha)}
+\sum_{J\in N(I)}{D_{J} \delta{f}_{J}^{(\alpha)}}\right)},
\end{equation}
where $\delta{f}_I^{(\alpha)} = f_I^{(\alpha)}-f_I^{(\alpha-1)}$ is the correction of the distribution function
obtained from each smoothing process.

Taking sufficient times in the smoothing process,
the residual equations (\ref{eq:smoothingMacro}) and (\ref{eq:rewrittenMicroDefect}) are supposed to be solved
to give the coarse-grid corrections.
In prediction step, the corrections of the conservative variables are obtained from $\Delta \vec{W}_H^{n} = \vec{W}_H^{n+1} - \vec{W}_H^{n}$
while in evolution step the total correction of the distribution function is computed by a summation of the intermediate corrections $\delta{f}^{(\alpha)}$ in each smoothing process.
Up to this point, we have obtained the corrections on the coarsest grid, which will be prolongated to finer grids
to reduce the low-frequency solution error on finer grids.

\subsubsection{Transfer operator : Prolongation}\label{sec:prolongation}
\begin{figure*}
	\centering
	\subfigure[\label{fig:innerCell}]{\includegraphics[width=0.32\textwidth]{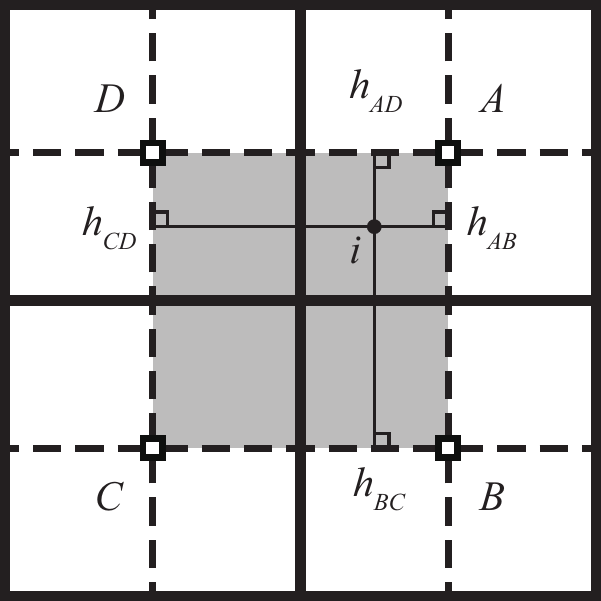}}
	\subfigure[\label{fig:boundaryCell}]{\includegraphics[width=0.32\textwidth]{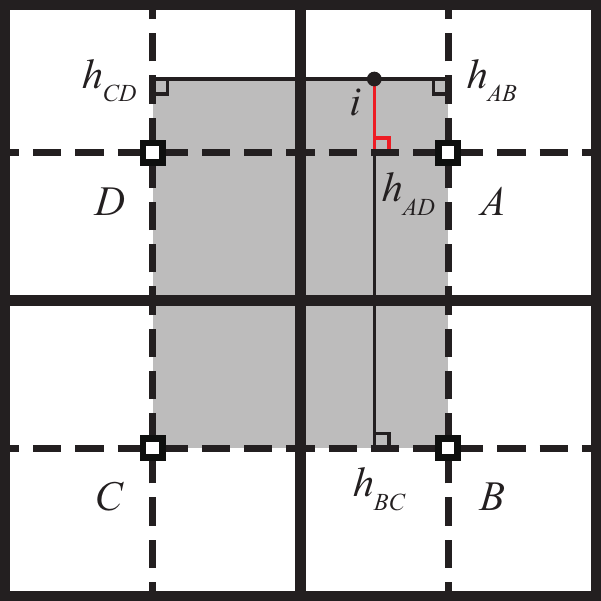}}
	\subfigure[\label{fig:cornerCell}]{\includegraphics[width=0.32\textwidth]{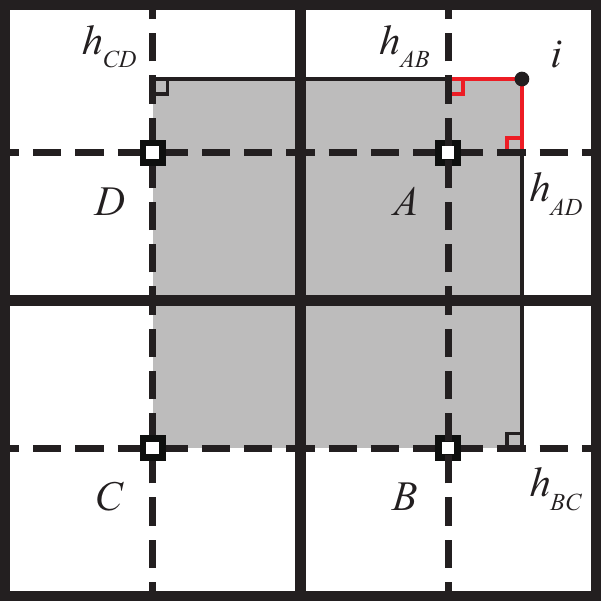}}
	\caption{\label{fig:prolongation}Stencils of prolongations for (a) inner cells, (b) boundary cells and (c) corner cells.}
\end{figure*}

The bilinear interpolation is used to prolongate the corrections from the coarser grids to finer ones. 
As shown in Fig.~\ref{fig:innerCell}, 
the interpolated result of a specific variable $Q_H$ on fine grid is
\begin{equation}\label{eq:prolongation}
(I_H^h Q_H)_i = \dfrac{\sum_{J\in S(i)}{w_J Q_J}}{\sum_{J \in S(i)}{w_J}},
\end{equation}
where $S(i)$ is the set of the coarse-grid stencil cells for the fine-grid cell $i$. The weights $w_J$ are
\begin{equation}\label{eq:prolongationWeight}
\begin{aligned}
w_A &= \dfrac{h_{BC} h_{CD}}{S_{ABCD}},\qquad
&w_B = \dfrac{h_{CD} h_{AD}}{S_{ABCD}},\\
w_C &= \dfrac{h_{AD} h_{AB}}{S_{ABCD}},\qquad
&w_D = \dfrac{h_{AB} h_{BC}}{S_{ABCD}},
\end{aligned}
\end{equation}
where $S_{ABCD} = (h_{AB}+h_{CD})(h_{BC}+h_{AD})$, and $h$ is the distance between the center of fine-grid cell $i$
and the line which connects the cell centers of two neighboring members in $S(i)$.

For instance, the prolongated correction of distribution function gives
\begin{equation}
(\Delta \bar{f}_h^n)_i = (I_H^h \Delta f_H^n)_i = w_A \Delta f_A^n +w_B \Delta f_B^n +w_C \Delta f_C^n +w_D \Delta f_D^n.
\end{equation}
Specifically, on Cartesian grids the weights give
\begin{equation}
\begin{aligned}
w_A &= \dfrac{9}{16},\qquad
&w_B = \dfrac{3}{16},\\
w_C &= \dfrac{1}{16},\qquad
&w_D = \dfrac{3}{16}.
\end{aligned}
\end{equation}

The weights given in Eq.~(\ref{eq:prolongationWeight}) are also available for extrapolation treatment of the cells near boundaries. In the cases of extrapolation some of the distance $h$ may be negative. For the cell $i$ near boundary shown in Fig.~\ref{fig:boundaryCell}, $h_{AD}$ is negative while for the corner cell illustrated in Fig.~\ref{fig:cornerCell} both $h_{AB}$ and $h_{AD}$ are negative. Specifically on Cartesian grids, the weights satisfy
\begin{equation}
\begin{aligned}
w_A &= \dfrac{15}{16},\qquad
&w_B = -\dfrac{3}{16},\\
w_C &= -\dfrac{1}{16},\qquad
&w_D = \dfrac{5}{16}.
\end{aligned}
\end{equation}
for boundary cells and
\begin{equation}
\begin{aligned}
w_A &= \dfrac{25}{16},\qquad
&w_B = -\dfrac{5}{16},\\
w_C &= \dfrac{1}{16},\qquad
&w_D = -\dfrac{5}{16}.
\end{aligned}
\end{equation}
for corner cells.

\subsection{Extension to multiple grids}\label{sec:multigrid}
Each component in the multigrid method of implicit UGKS has been described above. In this subsection, the multigrid algorithm will be constructed from recursion of the two-grid cycle.

First we define the multigrid cycle in prediction step as
\begin{equation}\label{eq:FASCYC}
\vec{W}_{l}^{n+1} ={\rm FAS} \left(l,\vec{W}_{l}^n,\vec{R}_l^n,\text{Eq.~(\ref{eq:smoothingMacro})},\nu_1,\nu_2\right),
\end{equation}
and the multigrid cycle in evolution step as
\begin{equation}\label{eq:CSCYC}
\Delta f_l^{n} = {\rm CS} \left(l,r_l^n,\text{Eq.~(\ref{eq:rewrittenMicroDefect})},\nu_1,\nu_2\right).
\end{equation}
where $l$ is the level index, $\nu_1$ and $\nu_2$ are the times of pre-smoothing and post-smoothing respectively.

In details, the recursive descriptions of an FAS cycle,  i.e., Eq.~(\ref{eq:FASCYC}), in prediction step are the following.
\begin{description}
\item[(a)] Pre-smoothing
	\begin{itemize}
	\item calculate the forcing function on this level of grid $\Omega_l$ by Eq.~(\ref{eq:forcingFunction}).
	\item get a better approximation $\hat{\vec{W}}_l^n$ and the updated residual $\hat{\vec{R}}_l^n$ by applying $\nu_1$ times of smoothing through two procedures, i.e.,
		\begin{itemize}
		\item get the intermediate approximation $\hat{\vec{W}}_l^{(m)}$ by solving Eq.~(\ref{eq:smoothingMacro}) with LU-SGS iterations.
		\item renew the residuals by Eq.~(\ref{eq:renewResiduals}) with forcing function.
		\end{itemize}	
	\end{itemize}
\item[(b)] Coarse grid correction
	\begin{itemize}
	\item get the initial approximate solution $\vec{W}_{l+1}^n$ and residual $\vec{R}_{l+1}^n$ by restricting $\hat{\vec{W}}_l^n$ and $\hat{\vec{R}}_l^n$ from the fine grid $\Omega_l$  to the coarser grid $\Omega_{l+1}$, see Eq.~(\ref{eq:macroRestriction})
	\item compute a new approximate solution $\vec{W}_{l+1}^{n+1}$ on the coarse grid $\Omega_{l+1}$, which may be one of the following two cases
		\begin{itemize}
		\item if $l+1=N_l$, solve the residual equations by sufficient times of smoothing, see Eqs.(\ref{eq:smoothingMacro}), (\ref{eq:forcingFunction}) and (\ref{eq:renewResiduals})
		\item if $l+1<N_l$, apply another FAS cycle on this level
		\begin{equation}
		\vec{W}_{l+1}^{n+1} ={\rm FAS} \left(l+1,\vec{W}_{l+1}^n,\vec{R}_{l+1}^n,\text{Eq.~(\ref{eq:smoothingMacro})},\nu_1,\nu_2\right),
		\end{equation}		
		\end{itemize}
	\item get the correction $\Delta \vec{W}_{l+1}^n$ from the difference $\vec{W}_{l+1}^{n+1}-\vec{W}_{l+1}^n$.
	\item interpolate the corrections to the finer grid $\Omega_{l}$ obtaining $\Delta \bar{\vec{W}}_{l}^n$ by Eq.~(\ref{eq:prolongation}).
	\item update the solution to $\bar{\vec{W}}_l^{n}$ on $\Omega_l$ 
\end{itemize}
\item[(c)] Post-smoothing
	\begin{itemize}
	\item  get a smoothed approximate solution $\vec{W}_l^{n+1}$ by applying $\nu_2$ steps of smoothing, with the following two steps:
		\begin{itemize}
		\item update the residual with the approximate solution and forcing function by Eq.~(\ref{eq:renewResiduals}).
		\item get the smoothed solution by solving Eq.~(\ref{eq:smoothingMacro}).
		\end{itemize}
	\end{itemize}
\end{description}

The recursive CS cycles in Eq.~(\ref{eq:CSCYC}) for the evolution step can be described as follows.
\begin{description}
\item[(a)] Pre-smoothing
	\begin{itemize}
	\item get the approximate correction $\Delta \hat{f}_l^n$ and update the residual to $\hat{r}_l^n$ by applying $\nu_1$ times of smoothing with two procedures, i.e.
		\begin{itemize}
		\item get the intermediate correction $\Delta \hat{f}_l^{(m)}$ by solving Eq.~(\ref{eq:rewrittenMicroDefect}) with LU-SGS iterations.
		\item renew the residuals by Eq.~(\ref{eq:renewResidualMicro}).
		\end{itemize}	
	\end{itemize}	
\item[(b)] Coarse grid correction
	\begin{itemize}
	\item get the residual ${r}_{l+1}^n$ by restricting $\hat{r}_l^n$ from the fine grid $\Omega_l$  to the coarser grid $\Omega_{l+1}$ by Eq.~(\ref{eq:microRestriction})
	\item compute a new approximate correction $\Delta {f}_{l+1}^{n}$ on the coarse grid $\Omega_{l+1}$, which may be one of the following two cases
		\begin{itemize}
		\item if $l+1=N_l$, solve the residual equation by sufficient times of smoothing, see Eqs.(\ref{eq:rewrittenMicroDefect}) and (\ref{eq:renewResidualMicro}).
		\item if $l+1<N_l$, apply another one CS cycle on this level
		\begin{equation}
		\Delta {f}_{l+1}^{n} ={\rm CS} \left(l+1,{r}_{l+1}^n,\text{Eq.~(\ref{eq:rewrittenMicroDefect})},\nu_1,\nu_2\right),
		\end{equation}		
		\end{itemize}
	\item interpolate the correction back to finer grid $\Omega_{l}$ by Eq.~(\ref{eq:prolongation}) obtaining $\Delta \bar{f}_{l}^n$.
	\item update the total correction $\Delta \bar{f}_l^n + \Delta \hat{f}_l^n$ on $\Omega_l$. 
\end{itemize}
\item[(c)] Post-smoothing
	\begin{itemize}
	\item  get a smoothed correction $\Delta {f}_l^{n}$ by applying $\nu_2$ steps of smoothing, similarly following two steps:
		\begin{itemize}
		\item update the residual by Eq.~(\ref{eq:renewResidualMicro}).
		\item get the smoothed correction by solving Eq.~(\ref{eq:rewrittenMicroDefect}).
		\end{itemize}
	\end{itemize}
\end{description}

With these two recursive definition of multigrid cycles of prediction step and evolution step, the implicit UGKS on multiple grids can be described as
\begin{description}
\item[Step 1\label{item:start}] calculate the time-averaged fluxes over physical time step and get the residual of conservative variables by Eq.~(\ref{eq:macroResidualDef});
\item[Step 2] obtain the predicted conservative variables from
\begin{equation}
\tilde{\vec{W}}_{l=0}^{n+1} = {\rm FAS} \left(l=0,\vec{W}_{l=0}^n,\vec{R}_{l=0}^n,\text{Eq.~(\ref{eq:smoothingMacro})},\nu_1,\nu_2\right);
\end{equation}
\item[Step 3] obtain the predicted equilibrium state $\tilde{g}_{l=0}^{n+1}$ from $\tilde{\vec{W}}_{l=0}^{n+1}$;
\item[Step 4] get the residual $r_{l=0}^n$ on the finest grid from Eq.~(\ref{eq:microResidualDef}) with predicted equilibrium state;
\item[Step 5] obtain the total correction of distribution function $\Delta f_{l=0}^{n}$ from
\begin{equation}
\Delta {f}_{l=0}^{n} ={\rm CS} \left(l=0,{r}_{l=0}^n,\text{Eq.~(\ref{eq:rewrittenMicroDefect})},\nu_1,\nu_2\right);
\end{equation}
\item[Step 6] get the updated the distribution function $f_{l=0}^{n+1}$ by the total correction;
\item[Step 7] update the conservative variables $\vec{W}_{l=0}^{n+1}$ and equilibrium state $g_{l=0}^{n+1}$ from compatible condition;
\item[Step 8] check the residual of conservative variables
\begin{itemize}
\item if the convergent state is reached, stop the calculation,
\item if not, go to  \nameref{item:start}.
\end{itemize}
\end{description}

\section{Remarks and discussions}\label{sec:remarks}
\subsection{Mesh generation}\label{sec:meshGeneration}
Different from the algebraic multigrid method (AMG) \cite{stuben2001review} which is based on mathematic treatment, the current multigrid method  adopts  geometric multigrid technique, so the concrete multiple grids should be generated first. For a given problem defined on a specific resolution (i.e., on a given discretized mesh), the multiple grids could be generated from the given discretization in physical space by coarsening algorithm level by level. Another way is to start with a coarsest grid to generate a satisfactory finest grid by refinement. In current paper the coarsening method from finest mesh is chosen to ensure that the convergent solution is defined on the original numerical discretization.

\begin{figure}
	\centering
	\includegraphics[width=0.96\columnwidth]{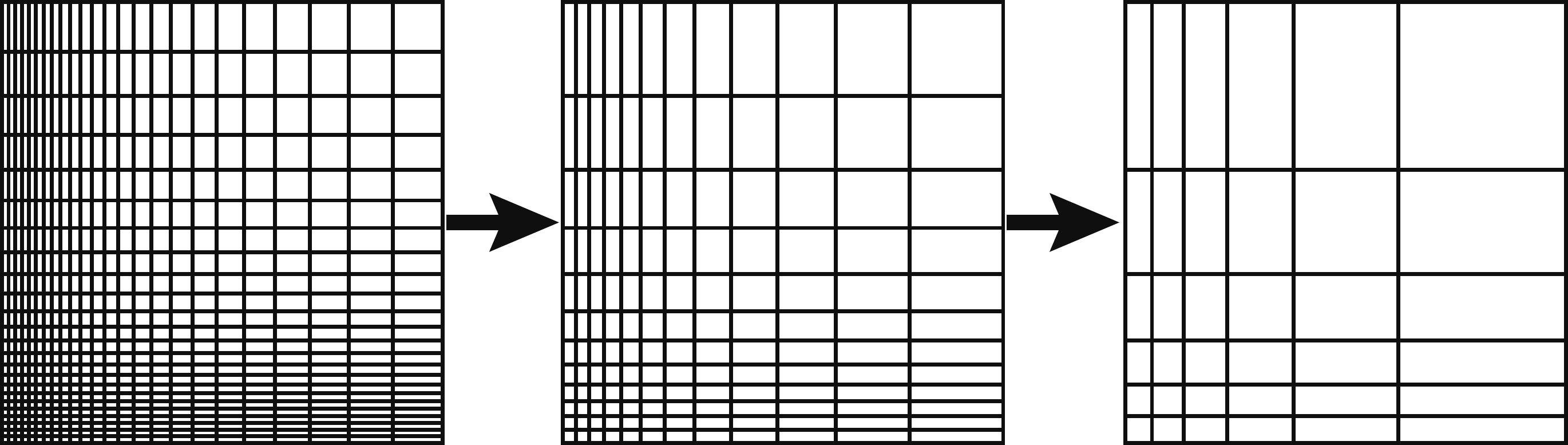}
	\caption{\label{fig:meshGeneration}Generation of triple level grids.}
\end{figure}

For structured grid, it is straightforward to generate the coarse meshes by deleting the grid points on every second line in each direction, see in Fig.~\ref{fig:meshGeneration}. 
After $N_l-1$ times of coarsening, $N_l$ levels of grids would be obtained. In this situation, the finest grid should satisfy
$$
N_p = 2^n+1, \quad n\ge N_l-1.
$$
where $N_p$ is the number of grid points in each direction. From Fig.~\ref{fig:meshGeneration}, it can be observed
that the coarsening method can retain more information of the finest grid, e.g., the growth rate of cell size along the x- and y-directions.
For unstructured mesh, generation methods, such as nonnested grids, topological methods and agglomeration methods, were introduced in \cite{blazek2001book,mavriplis1995multigrid}, about which we will not further discuss in current paper.

\subsection{Boundary condition}
Generally, boundary conditions in explicit schemes can be imposed by employing ghost cells. To solve the corrections from residual equations, the quantities in the delta-form in ghost cells should be given to start the sweeps for LU-SGS iterations. As described in \cite{zhu2016implicit},
the boundary conditions for the implicit scheme on each level of grid are derived from those in the explicit UGKS.

For the conservative flow variables, the relation between the ghost cell $j$ and the inner cell $i$ can be expressed in the following form
\begin{equation}\label{eq:boundaryRelation}
{\vec W}_j = {\vec B}\left({\vec W}_i\right),
\end{equation}
where ${\vec B}$ represents a specific transformation relation.
The linearization of the above equation gives
\begin{equation}\label{eq:boundaryMacro}
{\Delta {\vec W}_j^{n+1}} - \left(\dfrac{\partial {\vec B}}{\partial {\vec W}}\right)_i^{n} {\Delta {\vec W}_i^{n+1}} = {\vec 0},
\end{equation}
which is the macroscopic governing equation of the boundary conditions adopted in the smoothing process of the prediction step.
Here, we give the boundary condition for the isothermal walls to illustrate the treatment.
For a solid wall moving velocity ${\vec U}_w = (U_w,V_w)$ at temperature of $T_w$, the macroscopic variables in the ghost cells are
\begin{equation}\label{eq:boundaryWall}
\begin{aligned}
\rho_j &= \rho_i ,\\
\rho_j U_j    &= 2 \rho_i U_w - \rho_i U_i ,\\
\rho_j V_j   &= 2 \rho_i V_w - \rho_i U_i ,\\
\lambda_j &= 2 \lambda_w - \lambda_i,
\end{aligned}
\end{equation}
where $\lambda = 1/2 R T$.
After linearization, the changes of the conservative variables in the ghost cell vary with the values in the inner cell by
\begin{equation}\label{eq:example_results}
\left[
\renewcommand {\arraystretch}{1.2}
\begin{array}{*{20}{c}}
\Delta \rho_j\\
\Delta (\rho U)_j\\
\Delta (\rho V)_j\\
\Delta (\rho \varepsilon)_j
\end{array}
\right]
=
\left[
\renewcommand {\arraystretch}{1.2}
\begin{array}{*{20}{c}}
1 & 0 & 0 & 0\\
2 U_0 & -1 & 0 & 0\\
2 V_0 & 0 & -1 & 0 \\
\dfrac{\partial (\rho \varepsilon)_j}{\partial \rho_i} & \dfrac{\partial (\rho \varepsilon)_j}{\partial (\rho U)_i} &  \dfrac{\partial (\rho \varepsilon)_j}{\partial (\rho V)_i} & \dfrac{\partial (\rho \varepsilon)_j}{\partial (\rho \varepsilon)_i}
\end{array}
\right]
\left[
\renewcommand {\arraystretch}{1.2}
\begin{array}{*{20}{c}}
\Delta \rho_i\\
\Delta (\rho U)_i\\
\Delta (\rho V)_i\\
\Delta (\rho \varepsilon)_i
\end{array}
\right],
\end{equation}
where $(\rho, \rho U, \rho V, \rho \varepsilon)$ are the conservative variables. And we have
\begin{displaymath}
\begin{aligned}
\dfrac{\partial (\rho \varepsilon)_j}{\partial \rho_i}  &= 2 (U_w^2+V_w^2)-\frac{1}{2}(U_i^2+V_i^2)(1+\Upsilon^2)+\dfrac{K}{4 \lambda_i }(\Upsilon^2 - \Upsilon)\\
\dfrac{\partial (\rho \varepsilon)_j}{\partial (\rho U)_i}    &= U_i - 2 U_w + U_i \Upsilon^2 ,\\
\dfrac{\partial (\rho \varepsilon)_j}{\partial (\rho V)_i}    &= V_i - 2 V_w + V_i \Upsilon^2 ,\\
\dfrac{\partial (\rho \varepsilon)_j}{\partial (\rho \varepsilon)_i} &= - \Upsilon^2
\end{aligned}
\end{displaymath}
where $K$ is the total dimensions of degree of freedom and $\Upsilon = \lambda_i/(\lambda_i - 2 \lambda_w)$.
Similarly, the boundary conditions for gas distribution function in the smoothing process of the evolution step can be derived from
\begin{equation}
f_{j,k} = B(f_{i}).
\end{equation}
We have
\begin{equation}
\Delta f_{j,k} = (\partial B / \partial f_i) \Delta f_{i,k^\prime},
\end{equation}
where $f_{i,k^\prime}$ is the distribution function at velocity $\vec{u}_{k^\prime}$ corresponding to that in the ghost cells at velocity $\vec{u}_k$.
For horizontal symmetric interfaces, we have $\Delta f_{j,k} = \Delta f_{i,k^\prime}$ for $u_{k,x} = u_{k^\prime,x}$ and $u_{k,y} = u_{k^\prime,y}$. For an isothermal solid wall with moving velocity ${\vec U} = (U_w,V_w)$ and temperature $T_w$,
for the diffusive reflection boundary condition the reduced distribution function in ghost cells is
\begin{equation}\label{eq:f_on_wall}
f_{j,k} = \rho_j  \dfrac{1}{2 \pi R T_w} {\rm {exp}}{\left[-\dfrac{1}{2 R T_w}\left( \left(u_k - U_w\right)^2 + \left(v_k-V_w\right)^2 \right) \right]} = \rho_j C_k ,
\end{equation}
where $C_k$ is a constant for each discrete velocity.
Therefore, the variation of gas distribution function in ghost cells is determined by
\begin{equation}\label{eq:delta_f_on_wall}
\Delta f_{j,k} = C_k {\Delta \rho}_j,
\end{equation}
where ${\Delta \rho}_j $ is computed by no-transmission condition
\begin{equation}
\begin{aligned}
&{\Delta \rho_j} \int_{u_n < 0} {u_n \dfrac{1}{2 \pi R T_w} {\rm {exp}}{\left[-\dfrac{1}{2 R T_w}\left( \left(u -U_w\right)^2 + \left(v-V_w\right)^2\right)\right]} dudv} \\
&= -\sum_{u_{n,k}>0}{u_{n,k} {\Delta f}_{i,k} w_k},
\end{aligned}
\end{equation}
where $w_k$ is the weight at velocity $\vec{u}_k$ for numerical integrations.

The boundary conditions should be imposed when the LU-SGS sweeps come to the boundary cells
and when the residuals need to be re-evaluated before interpolations.

\subsection{Full multigrid (FMG) method}\label{sec:FMG}
\begin{figure}
	\centering
	\includegraphics[width=0.4\textwidth]{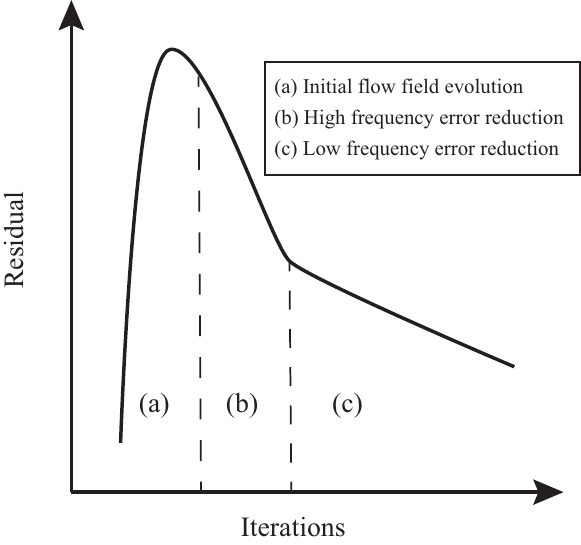}
	\caption{\label{fig:residualDiagram}Typical convergence characteristics for steady state solution with three stages, including (a) initial flow evolution stage, (b) high frequency error reduction stage, and (c) low frequency error reduction stage. }
\end{figure}

Generally, the evolution of residuals for a given case during calculations can be separated into three regions \cite{mavriplis1995multigrid} as shown in Fig.~\ref{fig:residualDiagram}.
In the first region, residuals will increase up to a peak value and start to decrease, representing the initial evolution of the flow field. In the second region, residuals decrease exponentially with iteration steps during which the high-frequency error is mainly eliminated. And for the last region, residuals continue decreasing but with a low efficiency because the low-frequency error needs more iterations to attenuate. The multigrid techniques speed up the second and third stages by eliminating  low-frequency error by enlarging the cell size. The full multigrid (FMG) method \cite{stuben1982multigrid,brandt2011guide} takes the first region into considerations as well to achieve a better efficiency.

\begin{figure*}
	\centering
	\subfigure[\label{fig:FMGcycle}]
	{\includegraphics[width=0.4656\textwidth]{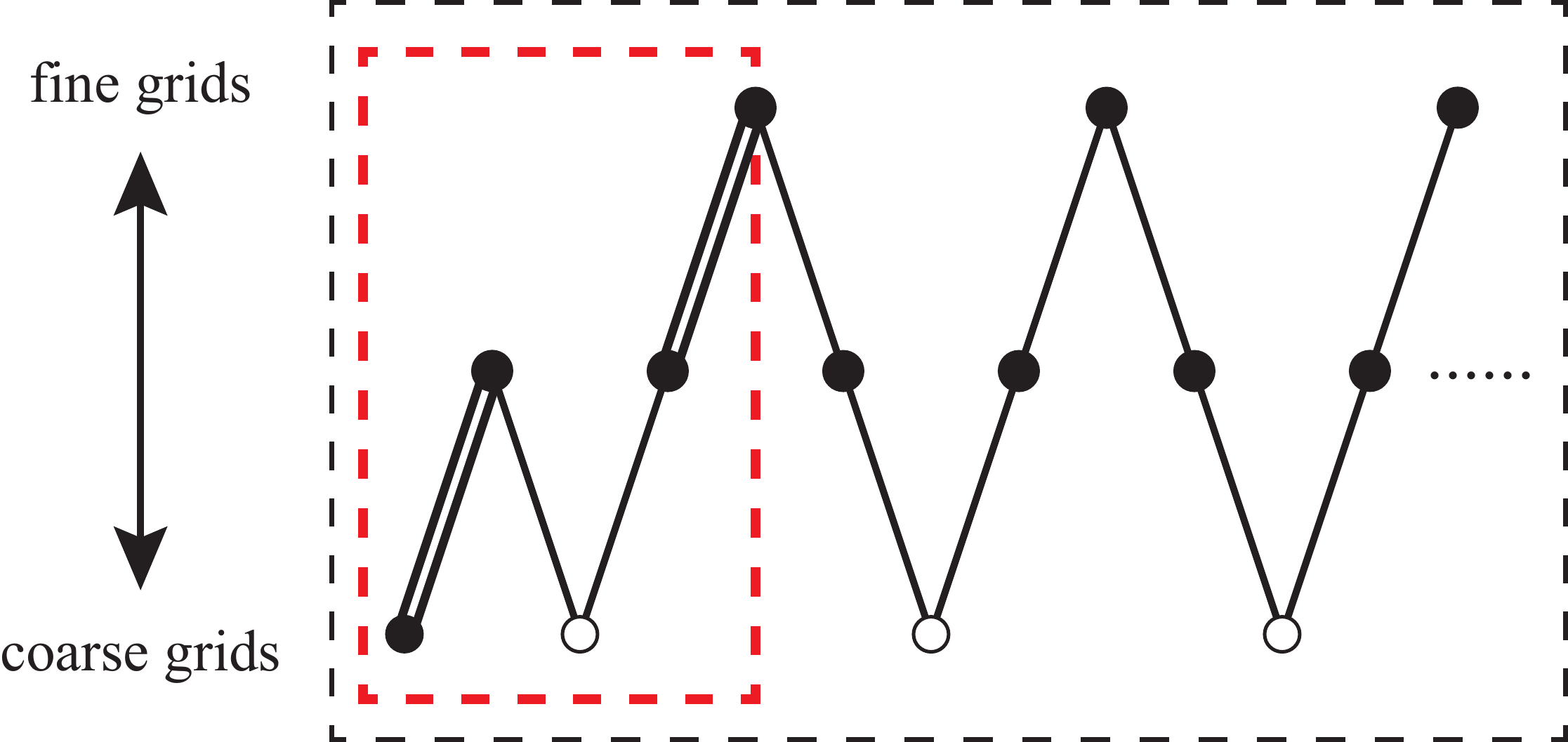}}\hspace{0.02\textwidth}
	\subfigure[\label{fig:Vcycle}]
	{\includegraphics[width=0.4\textwidth]{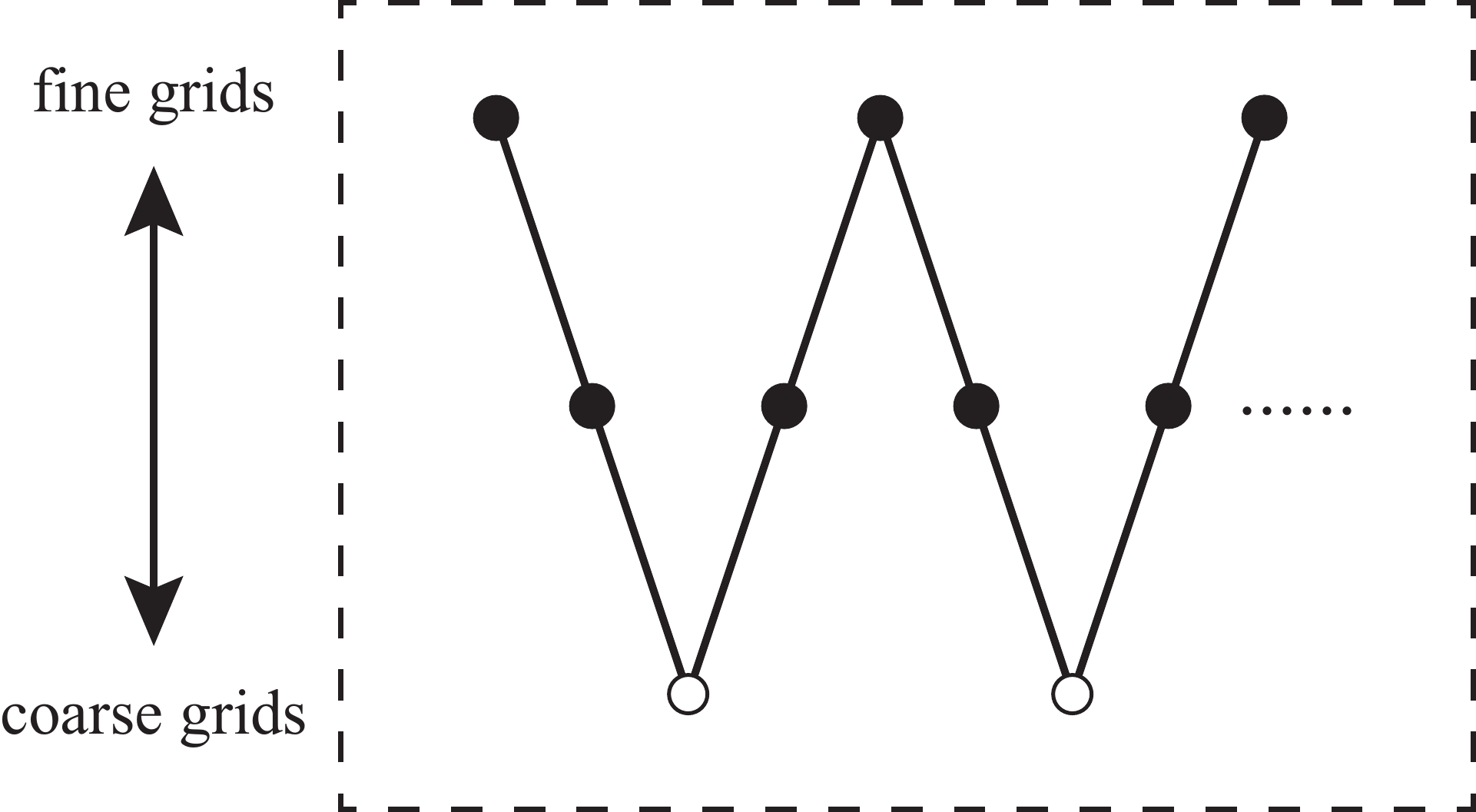}}
	\caption{\label{fig:FMG}Structure of (a) the full multigrid method in comparison with (b) the typical V-type cycle. Black circle: smoothing process; white circle: solving of residual equations; double slash: FMG interpolation; backslash: restriction; slash: prolongation.}
\end{figure*}

The FMG method starts from the coarsest level of grid to provide the initial approximate solutions for finer grids, which can reduce the evolution time of the flow field due to the lower computations on coarse grids. The structure of the FMG method in comparison with the V-type cycle is illustrated in Fig.~\ref{fig:FMG}.
The flow field will evolve on the coarser grids before being interpolated onto finer grids. Different from prolongation in multigrid cycles, the FMG interpolation denoted by double slash in Fig.~\ref{fig:FMGcycle} transfers the conservative variables and gas distribution functions from coarse grids to fine grids instead of their corrections, because it requires all flow variables to initialize the flow field on finer grids. It can be seen that the FMG method will be more effective in complex flows, for which more CPU time will be taken to evolve the initial flow fields using the explicit and implicit schemes on a single grid. With the similar idea, the algorithms with low order of accuracy can be also adopted to get a fully evolved approximate solution before taking a higher-order scheme.

\subsection{Miscellaneous factors}
There are many factors that should be considered in multigrid method to ensure the stability and convergence efficiency,
such as the type of cycles, number of smoothing steps, the accuracy of the transfer operators, and the chosen of time steps.
In the following, a brief discussion about these factors will be given.

There are several types of cycles, such as V-type and W-type, that are commonly used in multigrid methods. All types of cycles can be derived from the basic cycle, i.e., the two-grid cycle. Through recursion of two-grid cycles, the most natural derivation is the V-type cycle. The V-cycle of three-grid methods can be regarded as using a deeper two-grid cycle to solve the residual equation on the coarser grid of the two-grid cycle. Other types of multigrid cycle, such as W-cycle, F-cycle and adaptive cycle, can be constructed by different combination of the two-grid cycles for a better convergence efficiency. In the current paper, the influence of the cycle types will not be further discussed. Generally, the V-cycle is fast enough, but for complex situations advanced types may get better convergence efficiency.

Another factor that will influence the convergence speed is the number of the pre-smoothing and the post-smoothing on each level of grid. Although more smoothing steps may bring better convergence in one multigrid cycle, it will be more efficient not to smooth the error too much but rather carrying out a few more multigrid cycles. As demonstrated in \cite{trottenberg2000multigrid}, common choices are $\nu_1+\nu_2 \leq 3$ in practice. In this paper, two pre-smoothing steps are carried out before the restriction of residuals and one post-smoothing step is carried out after the prolongation for an upwind spatial discretization\cite{blazek2001book}.

The restriction and prolongation should also satisfy certain accuracy requirements \cite{trottenberg2000multigrid,blazek2001book}, i.e.,
\begin{equation}
m_R + m_P > m_E,
\end{equation}
where $m_R$ and $m_P$ are the orders of the accuracy of the restriction and prolongation operators, respectively.
$m_E$ is the order of the numerical scheme. As given in Section \ref{sec:restriction} and \ref{sec:prolongation}, the restriction by using the volume weighted interpolation and the prolongation by using bilinear interpolation give $m_R = m_P = 2$, and the second-order UGKS gives $m_E=2$.

As described in paper \cite{zhu2016implicit}, there are two time steps in the implicit UGKS. The physical time step $\Delta t_p$, which is used to calculate the time-averaged fluxes in the evaluation of residuals, is related to the cell size through the CFL condition. The other one is a pseudo-time step, namely the numerical time step $\Delta t_n$, which is applied in the temporal discretization of the governing equations. For steady state solutions, the numerical time step is not a necessity, therefore it does not appear in the description of the current multigrid method. However, the numerical time step does help to improve the stability of the implicit schemes for tough numerical cases. If necessary, a term of $1/\Delta t_n$ could be added into the coefficients before the errors $\vec{E}_i^n$ and $e_{i,k}^n$ in Eq.~(\ref{eq:macroFinalDefect}) and Eq.~(\ref{eq:microFinalDefect}). For instance, a numerical time step which increases exponentially with iteration steps is used in the calculation of the hypersonic flow around the square cylinder in Section \ref{sec:square}.

\section{Rarefied flow studies}\label{sec:numericalTests}
In this section, the multigrid implicit UGKS will be used to study rarefied flow phenomena from low and high speed at various Knudsen numbers. All UGKS computations in this section are carried out on a single machine with a processor of Intel(R) Core(TM) i5-4570 CPU@3.2GHz, and no parallel technique is adopted here. The comparison in terms of accuracy and computational efficiency between MIUGKS and DSMC, whenever available, will be presented.

\subsection{Lid-driven cavity flow}\label{sec:cavity}
Simulations of lid-driven cavity flows are studied at different Knudsen numbers.
Following the previous work \cite{huang2012unified,john2011effects}, the gas in the cavity is argon with molecular mass $m_0 = 6.63\times10^{-26} kg$  and with an initial temperature $T_0 = 273K$. The cavity has a fixed wall temperature $T_w=273K$ and a moving lid at a constant velocity $U_w =50 m/s$.
The Knudsen number is defined as the ratio of mean free path to the length of cavity side wall. The dynamic viscosity is evaluated by $\mu=\mu_{0} (T/T_{0})^\omega$ where $\omega=0.81$. Cases at three different Knudsen numbers, i.e., $Kn=10, 1.0, 0.075$ have been tested.

In physical space, the computational domain is discretized with a mesh of $64\times64$ cells. In velocity space, $120\times120$, $100\times 100$ and $80\times 80$ discrete velocity points are used respectively for cases at $Kn=10, 1.0$ and $0.075$. In all three cases, the trapezoidal rule is used in the integration of the discretized distribution function to get macroscopic variables. The steady state is thought to be reached when the mean squared residuals of the conservative variables are reduced to a level being less than $1.0\times10^{-6}$,
where the residuals are computed by
\begin{equation}\label{eq:Residual_Def}
\vec{R}^n = \sqrt{\sum_{i=1}^{N_c}{\vec{R}_i^2}/N_c}
\end{equation}
which denotes the variation rate of the conservative variables. Here $N_c$ is the total number of discrete cells in the computational domain.

\begin{figure*}
	\centering
	\subfigure[]{\includegraphics[width=0.32\textwidth]{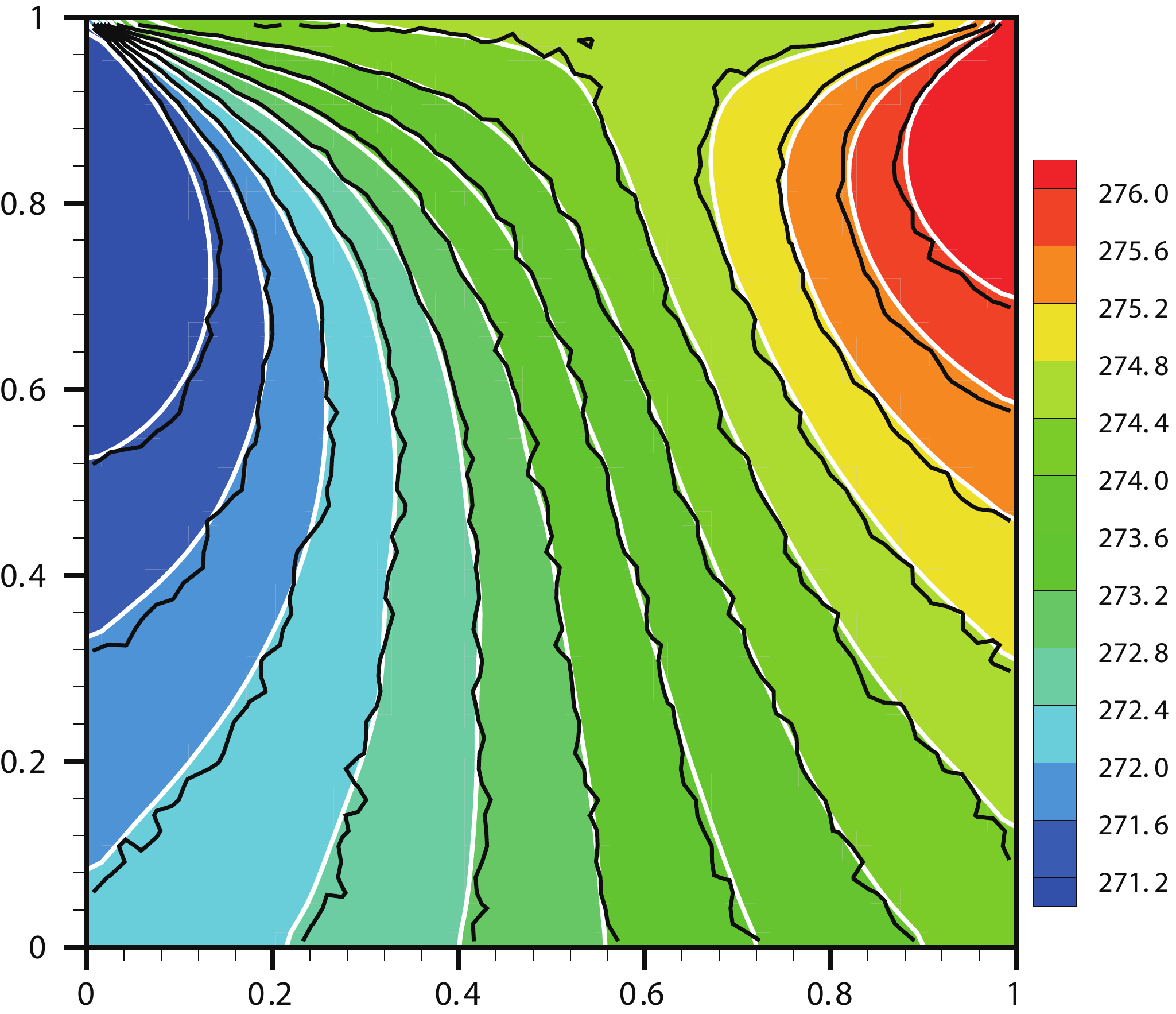}}
	\subfigure[]{\includegraphics[width=0.32\textwidth]{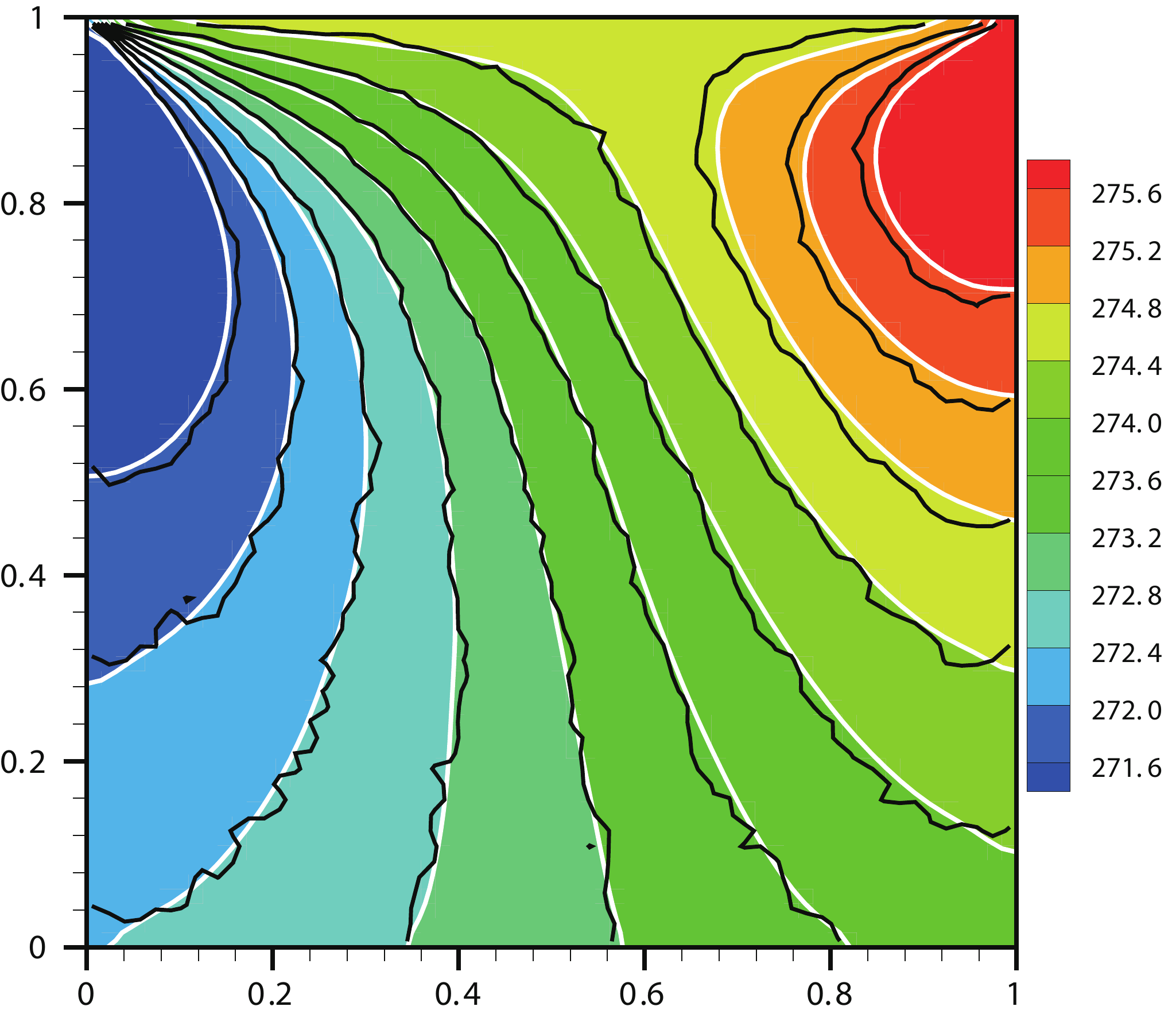}}
	\subfigure[]{\includegraphics[width=0.32\textwidth]{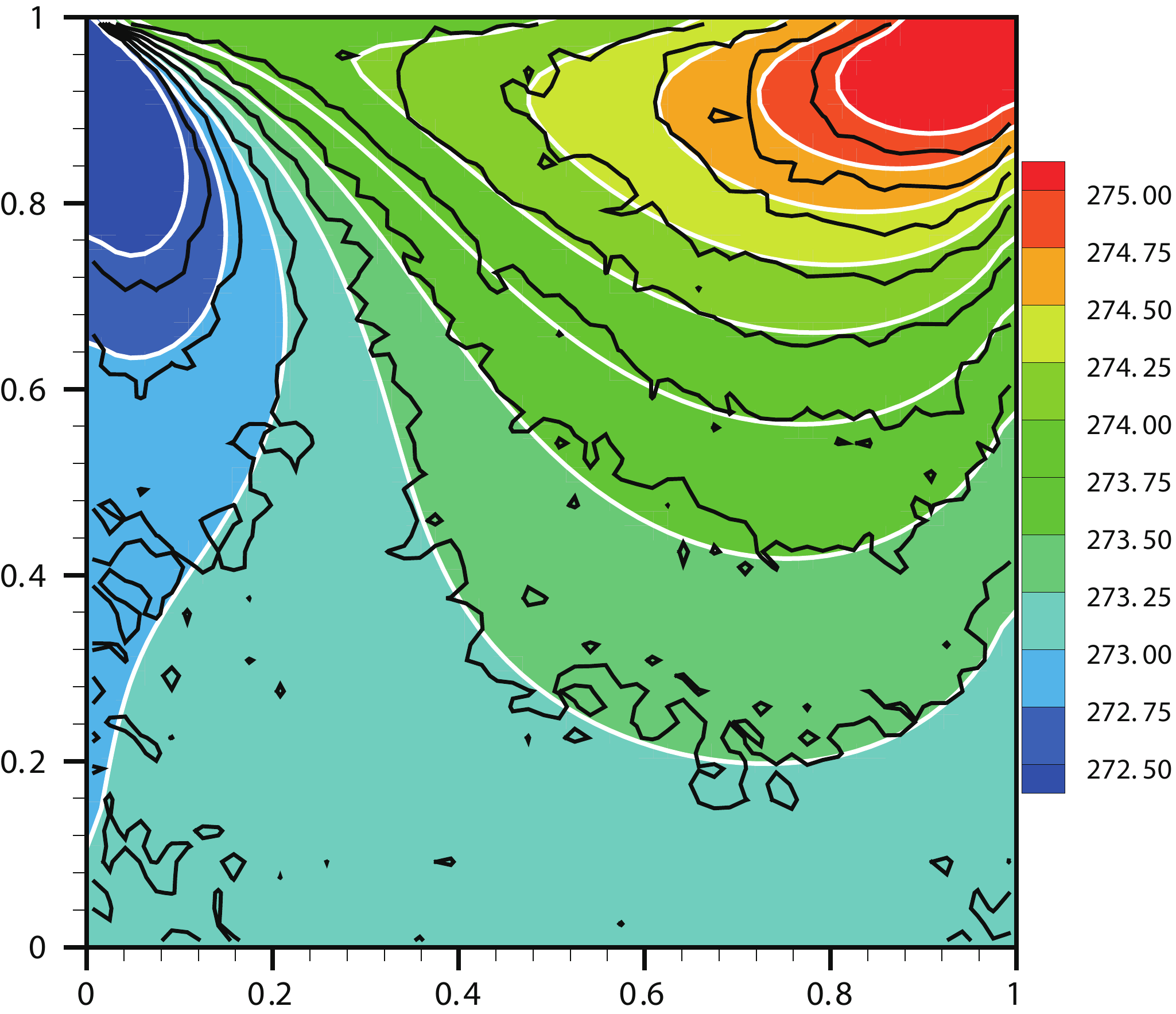}}
	\caption{\label{fig:cavityTemperature}Temperature distribution in the cavity at different Knudsen numbers: (a) $Kn=10$, (b) $Kn=1$ and (c) $Kn=0.075$. Background: the current multigrid method; white solid line: the implicit UGKS; black solid line: the DSMC results from paper \cite{huang2012unified}.}
\end{figure*}

\begin{figure*}
	\centering
	\subfigure[]{\includegraphics[width=0.32\textwidth]{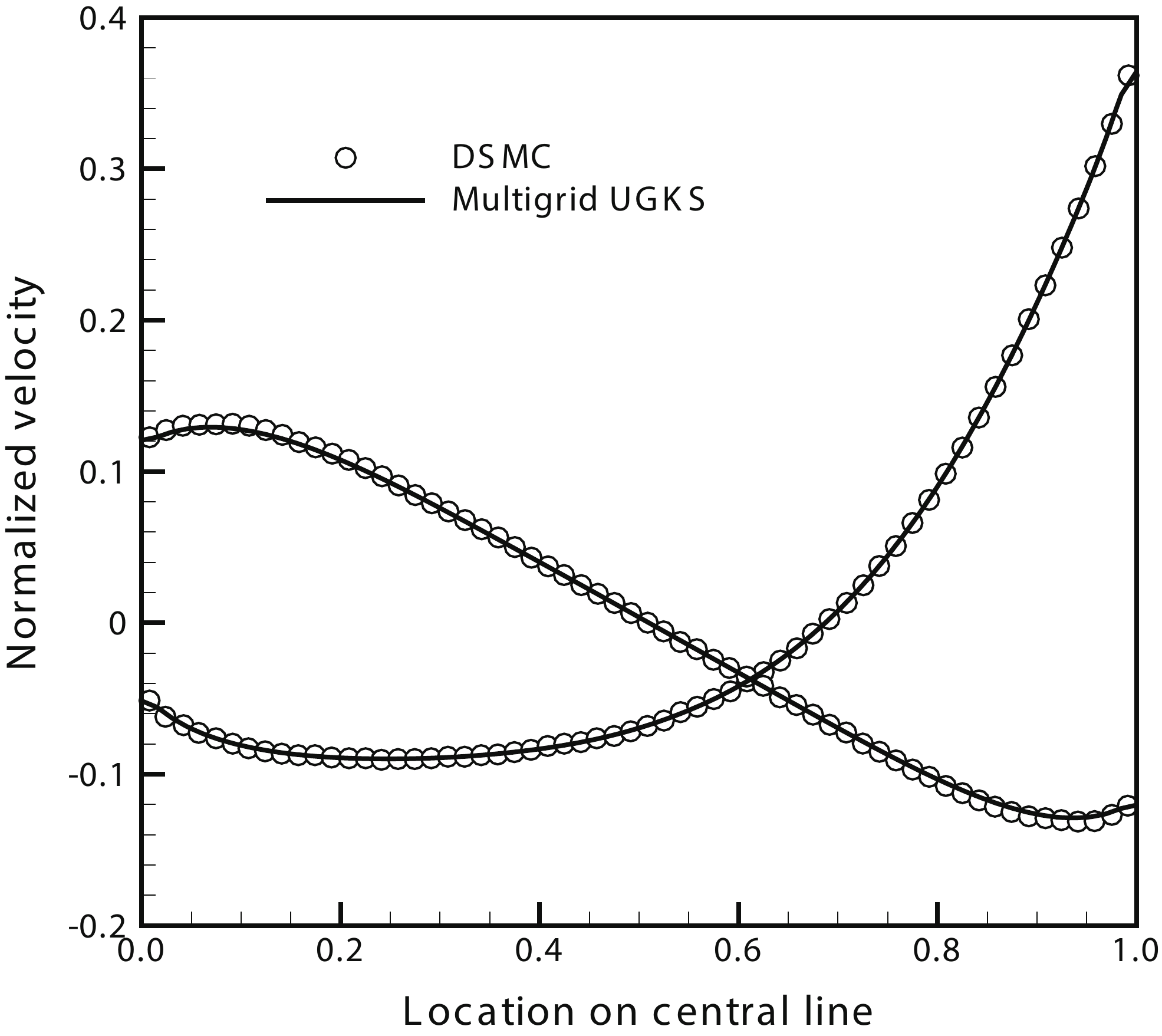}}
	\subfigure[]{\includegraphics[width=0.32\textwidth]{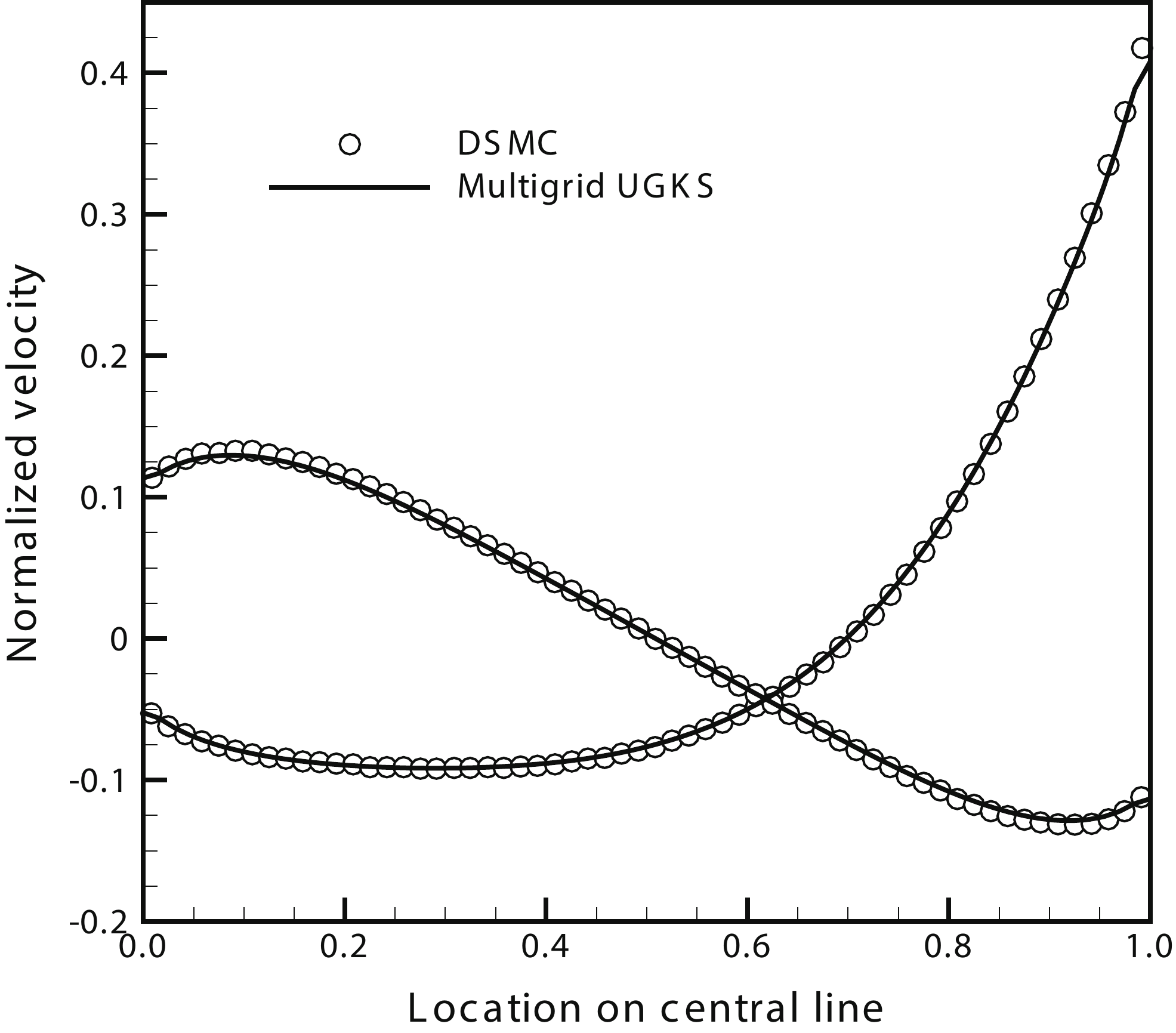}}
	\subfigure[]{\includegraphics[width=0.32\textwidth]{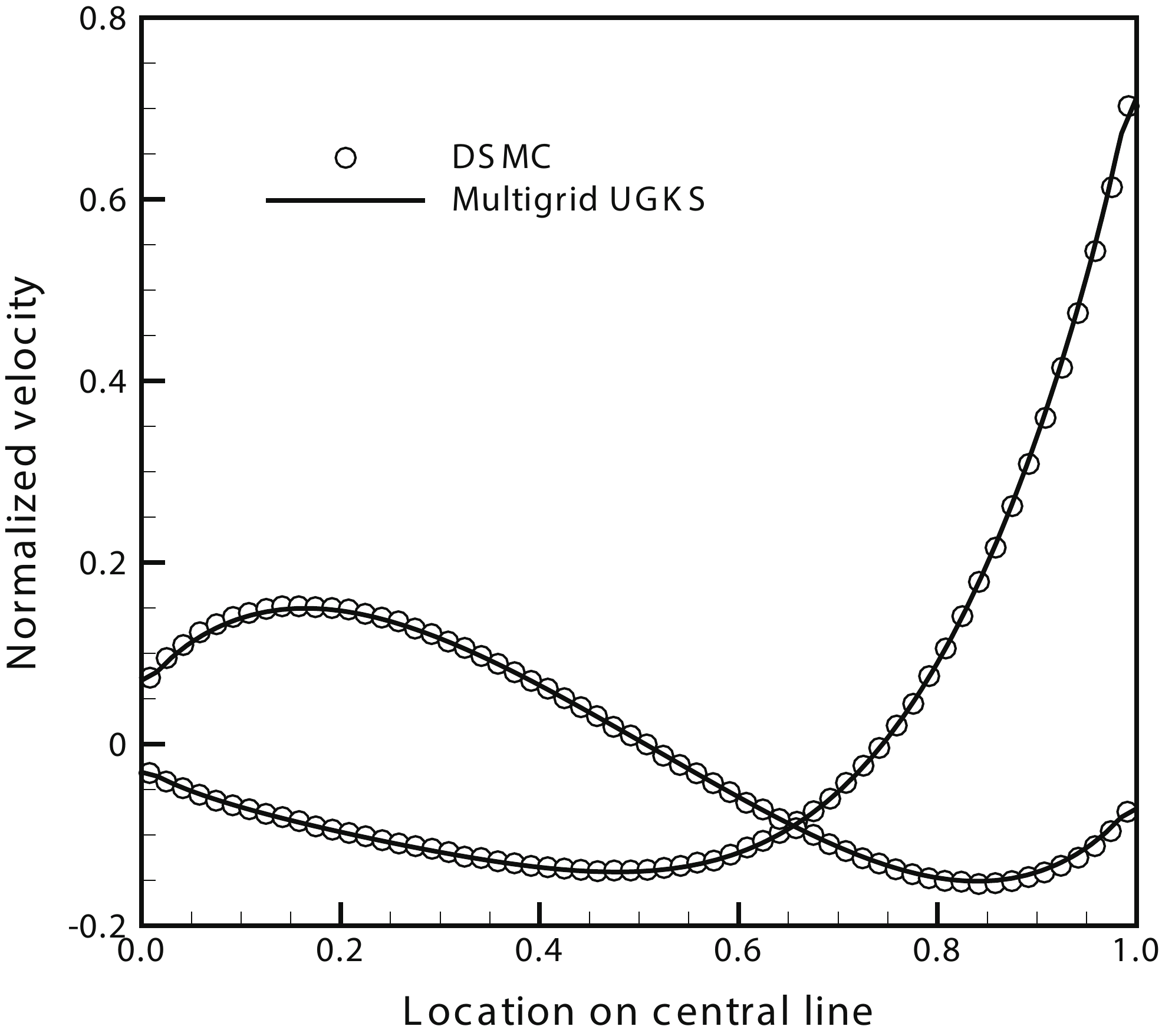}}
	\caption{\label{fig:cavityVelocity}Normalized X-velocity distribution along vertical central line and normalized Y-velocity distribution along horizontal central line at different Knudsen numbers: (a) $Kn=10$, (b) $Kn=1$ and (c) $Kn=0.075$.}
\end{figure*}

\begin{figure*}
	\centering
	\subfigure[]{\includegraphics[width=0.32\textwidth]{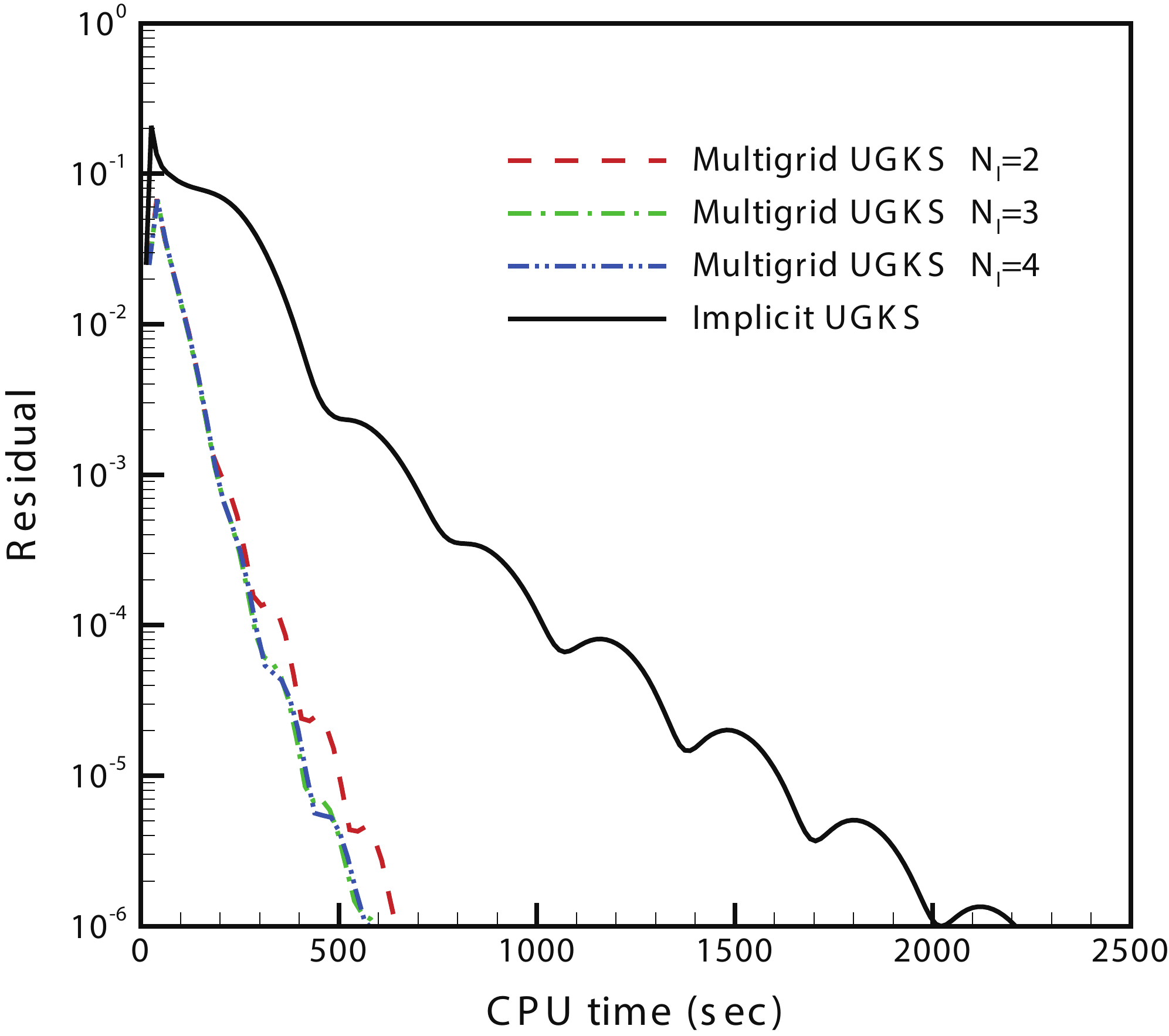}}
	\subfigure[]{\includegraphics[width=0.32\textwidth]{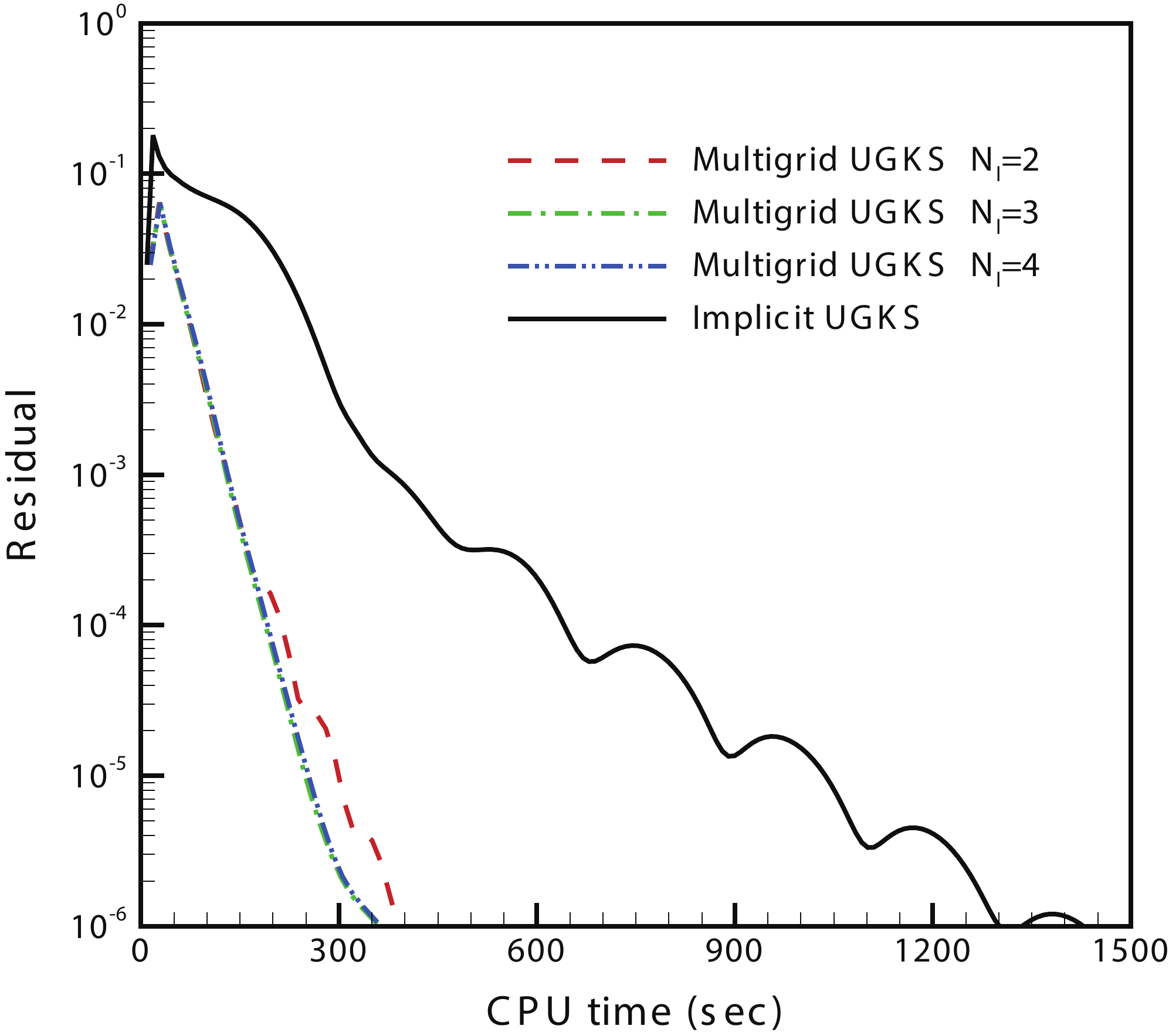}}
	\subfigure[]{\includegraphics[width=0.32\textwidth]{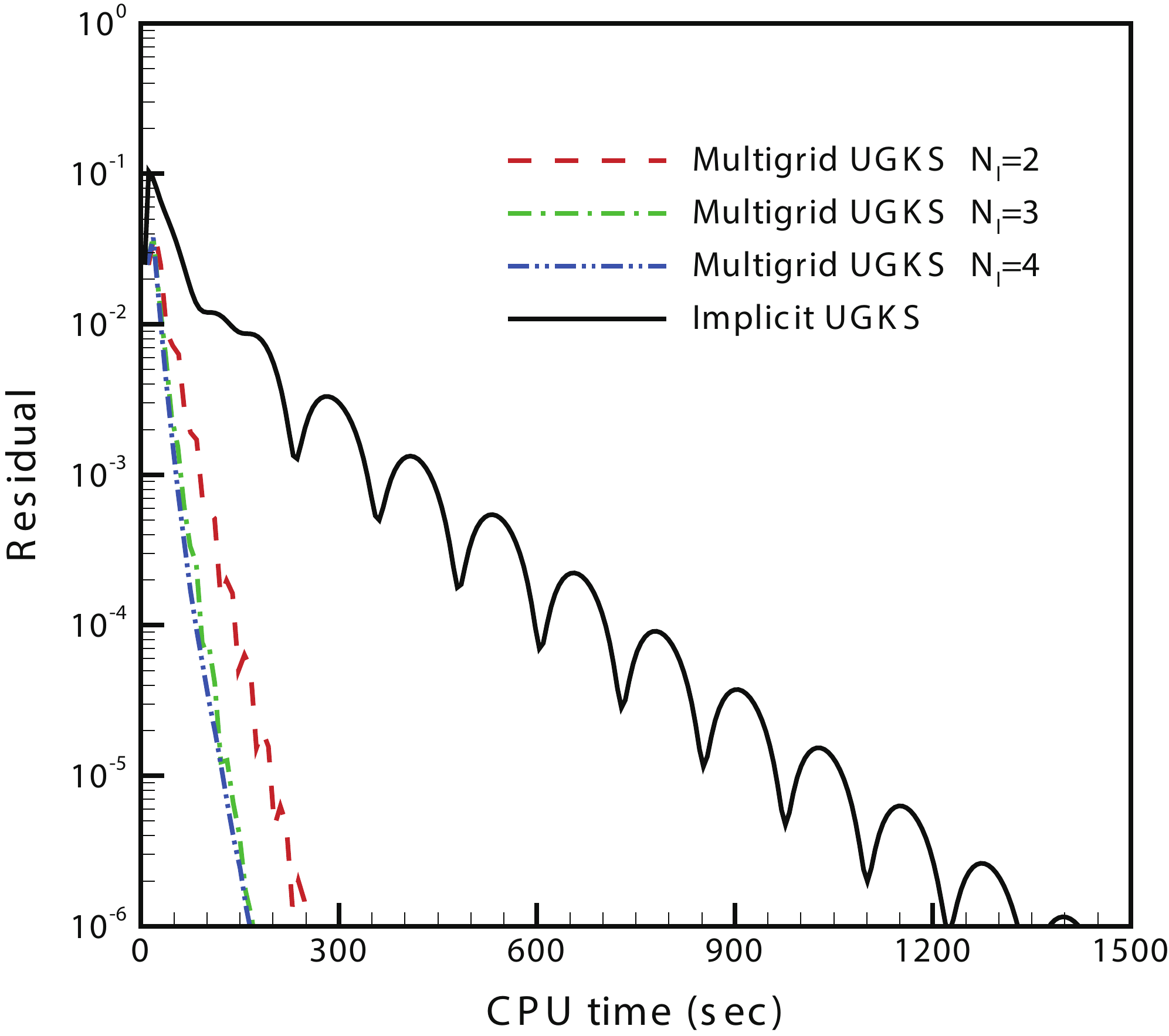}}
	\caption{\label{fig:cavityHistory}Convergence history of the cavity flow indicated by the residual of energy density at different Knudsen numbers: (a) $Kn=10$, (b) $Kn=1$ and (c) $Kn=0.075$. Black solid line: the implicit UGKS; red dashed line: two-level multigrid method; green dash-dotted line: three-level multigrid method; blue dash-dot-dotted line: four-level multigrid method.}
\end{figure*}

The results of the temperature distribution in the cavity at different Knudsen numbers have been plotted in Fig.~\ref{fig:cavityTemperature}. 
The results of the multigrid method are consistent with those of the original implicit UGKS with a single level of grid, and agree well with the DSMC results obtained from paper \cite{huang2012unified}. We also plot the distribution of the normalized velocities along the vertical and horizontal central lines comparing to the results of DSMC in Fig.~\ref{fig:cavityVelocity},
which also shows good agreement between the present results and reference data. In Fig.~\ref{fig:cavityHistory}, 
the convergence histories of the energy density with respect to CPU time are given to show the acceleration of the multigrid method. Obvious accelerating effects of the multigrid method on the implicit UGKS can be observed. Moreover, it can be seen that the multigrid methods with three-level grids and four-level grids converge faster than that with two-level grids. In comparison with the three-level grid method, the computation efficiency of the four-level scheme doesn't further increase because of the extra computations for another coarser level of grids. In the higher Knudsen number cases at $Kn=10$ and $1$, the multigrid method is about $3$ times faster than the original implicit scheme and in the case at $Kn=0.075$ the acceleration rate can be increased up to $8$ times. With a single machine (Intel(R) Core(TM) i5-4570 CPU@3.2GHz), at $Kn = 0.075$ the current scheme can get convergent solution with the CPU time being less than $3$ minutes, where the DSMC solution needs parallel supercomputers for that \cite{john2011effects}.

\subsection{Flat plate flow}\label{sec:plate}
Following the studies in paper \cite{sun2004drag}, subsonic flow passing over a flat plate with zero thickness at zero angle of attack is studied at different Reynolds numbers. The flat plate has a finite length with a fixed temperature of $T_w=295K$. The freestream is air with a temperature of $T_{\infty} = 295K$ at Mach number $0.2$. The dynamic viscosity coefficient is calculated by $\mu = \mu_{ref} (T/T_{ref})^\omega$ with a temperature model index $\omega=0.77$. The global Knudsen number and Reynolds number are defined with respect to the length of the plate and have a relation of $Kn\approx 1.19 Ma/Re$.

\begin{figure}
	\centering
	\includegraphics[width=0.6\columnwidth]{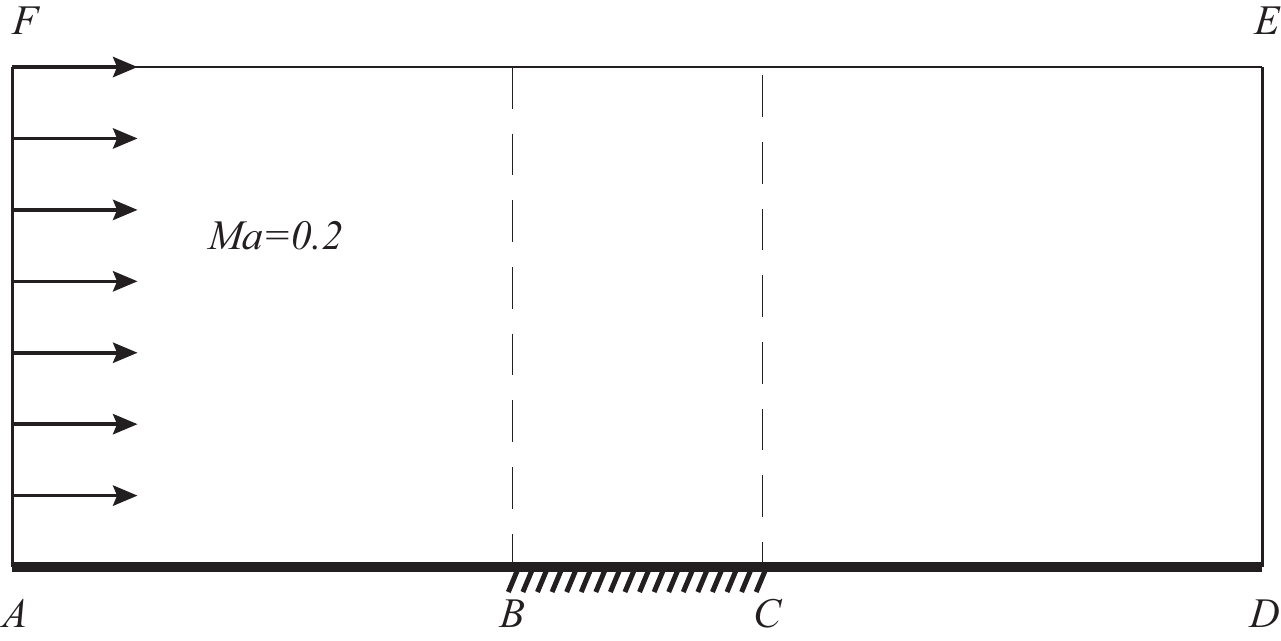}
	\caption{\label{fig:flatPlateSet} Illustration of the computational domain for the subsonic flow around flat plate.}
\end{figure}

Cases at different Reynolds numbers are studied using the current multigrid method and to explore its computation efficiency. In all cases, the farfield is 20 times of the plate length far away from the leading edge and trailing edge. As illustrated in Fig.~\ref{fig:flatPlateSet}, 
the distances satisfy $L_{AB}=L_{CD}=L_{AF}=20L_{BC}$. For each case, we adopt different discretization of the physical space and velocity space, see those listed in Table~\ref{tab:flatPlateDiscretization}.
\begin{table}
	\caption{\label{tab:flatPlateDiscretization}Discretization in physical space and velocity space for the flat plate flow simulation.}
	\begin{ruledtabular}	
		\begin{tabular}{cccccc}
			\multirow{2}{*}{Re} & \multicolumn{4}{c}{Physical space}  
			& \multirow{2}{*}{$N_u \times N_v$ \footnotemark[3]} \\ \cline{2-5} 
			~                                        & $N_{BC}$
			& $N_{AF}$ \footnotemark[1]              & $N_{\rm{total}}$ 
			& $\delta y _{\rm{min}}$ \footnotemark[2] & ~ \\ 
			\hline
			0.2                                      & 32 
			& 32                                     & 3072  
			& 0.02                                   & $100\times100$   \\
			0.5, 1                                   & 48 
			& 48                                     & 6912  
			& 0.01                                   & $90\times 90$    \\
			2, 5                                     & 48 
			& 48                                     & 6912 
			& 0.005                                  & $80\times 80$    \\
			10, 20, 50                               & 64 
			& 48                                     & 7680
			& 0.002                                  & $70\times 70$    \\
		\end{tabular}
	\end{ruledtabular}
	\footnotetext[1]{The numbers of discrete cells on edges AF, AB and 	CD are equal, i.e., $N_{AF}=N_{AB}=N_{CD}$.}
	\footnotetext[2]{$\delta y_{\rm{min}}$ denotes the minimum height of cells near the solid boundary.}
	\footnotetext[3]{The numbers of discrete velocity points in phase space.}
\end{table}
Symmetric boundary condition is used for the axes in the upstream and downstream, and diffusive reflection with full thermal accommodation coefficient is adopted in the boundary condition for the isothermal solid wall.

\begin{figure}
	\centering
	\subfigure[]{\includegraphics[width=0.96\columnwidth]{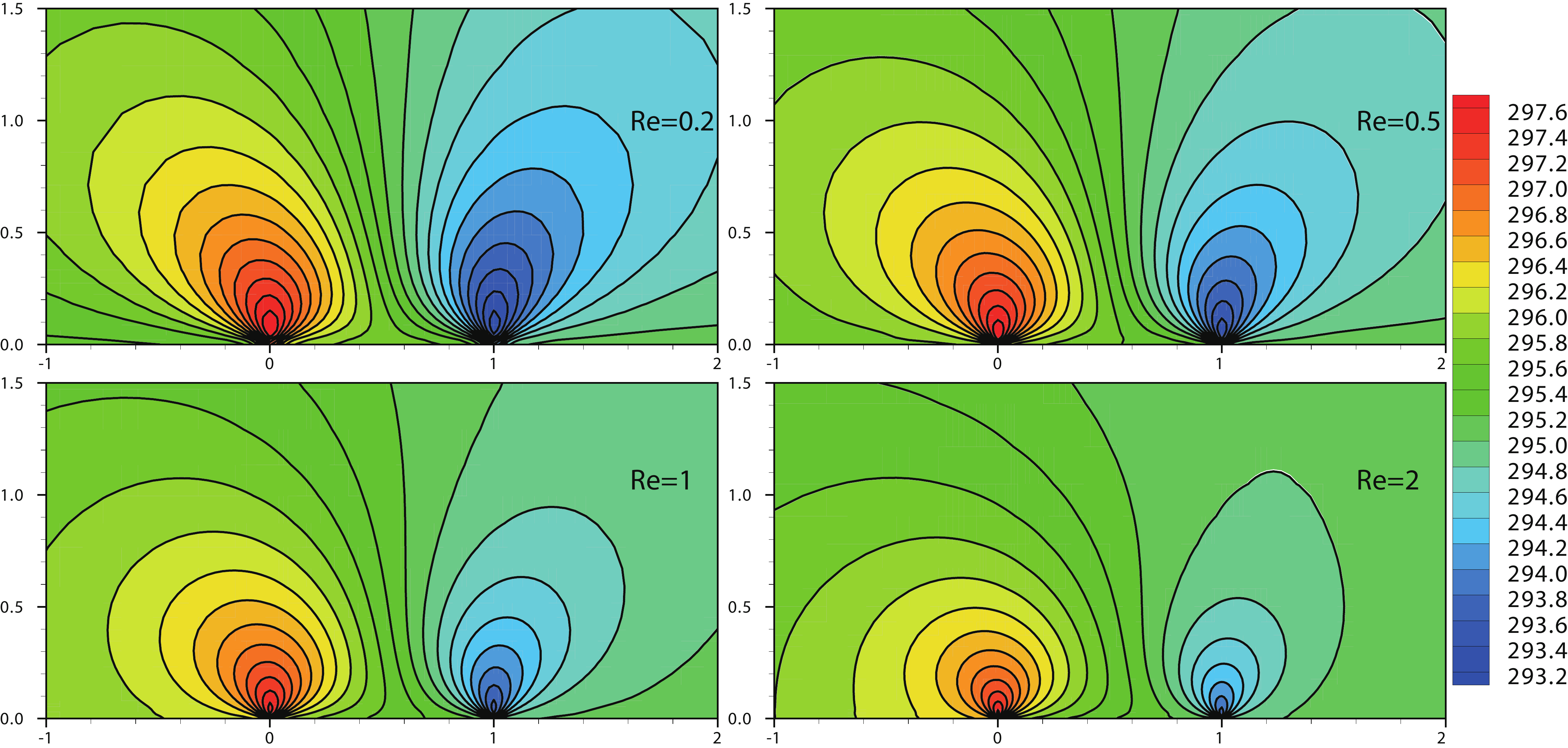}}\\
	\subfigure[]{\includegraphics[width=0.96\columnwidth]{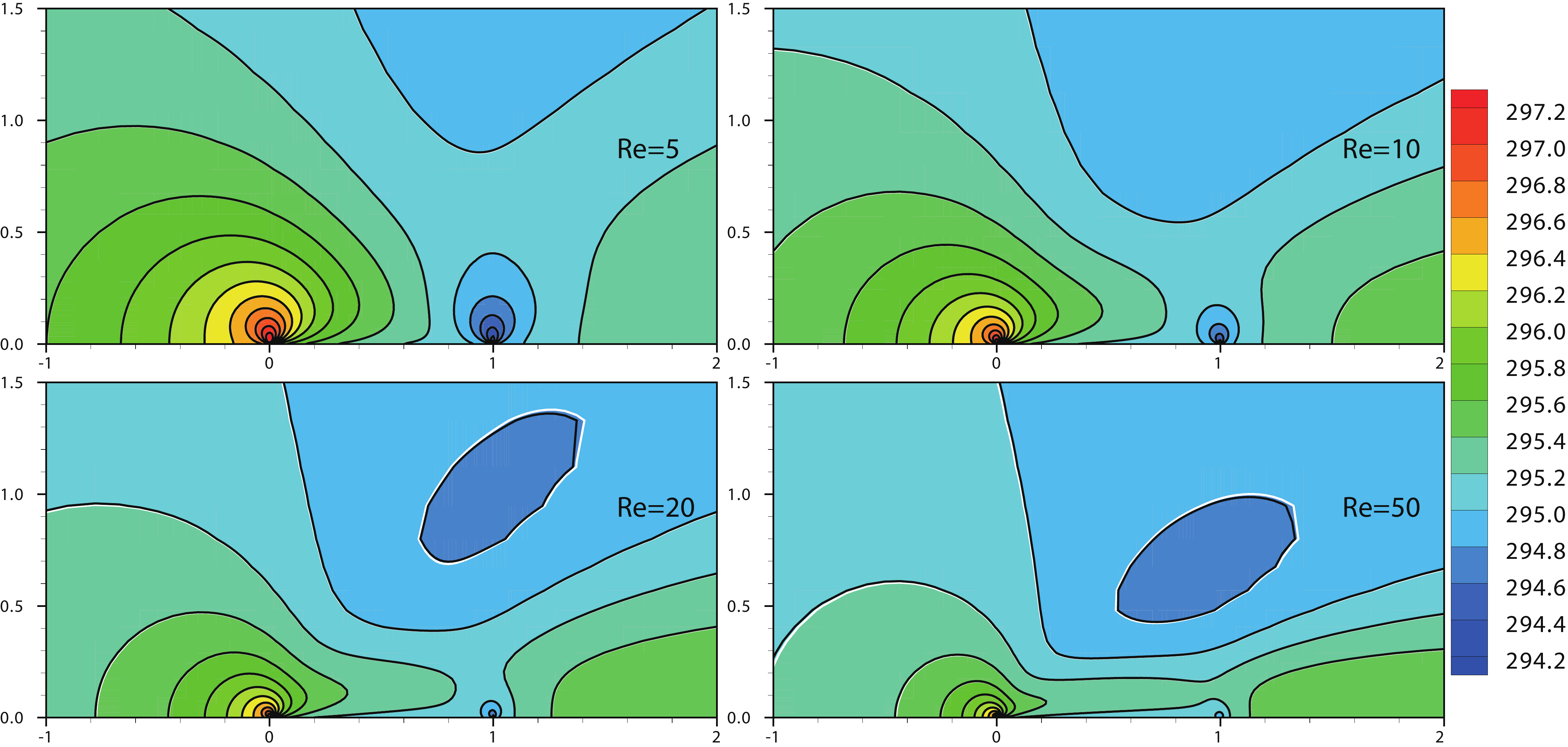}}
	\caption{\label{fig:flatPlateTemperature}The temperature distribution around the flat plate at different Reynolds numbers. The background with white line: the implicit UGKS solutions; the dashed line: the current multigrid method results.}
\end{figure}

\begin{figure}
	\centering
	\includegraphics[width=0.5\textwidth]{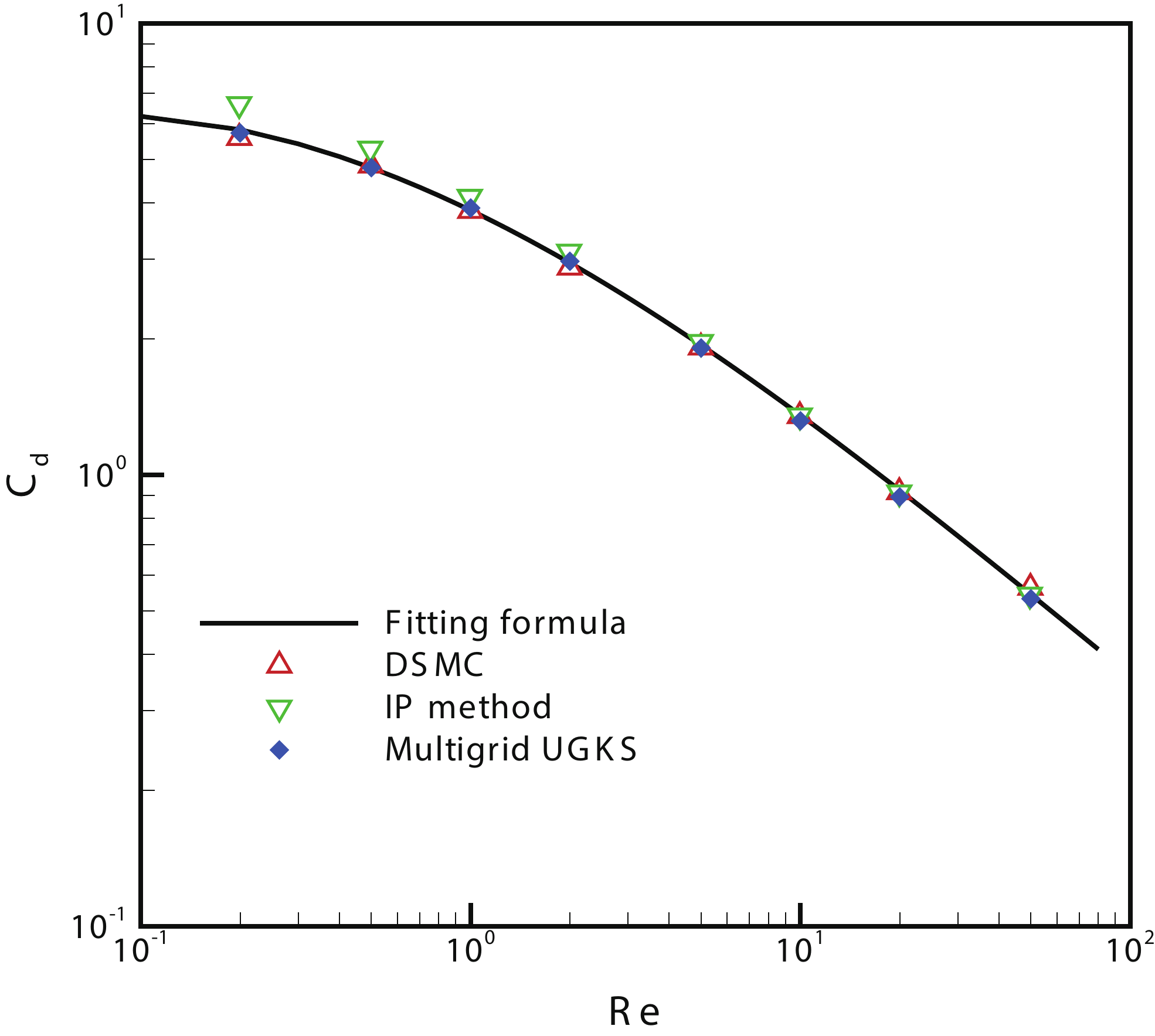}
	\caption{\label{fig:flatPlateDrag} Drag coefficients of the flat plate with respect to Reynolds number.}
\end{figure}

\begin{figure}
	\centering
	\subfigure[\label{fig:flatPlateFrictionA}]{\includegraphics[width=0.48\textwidth]{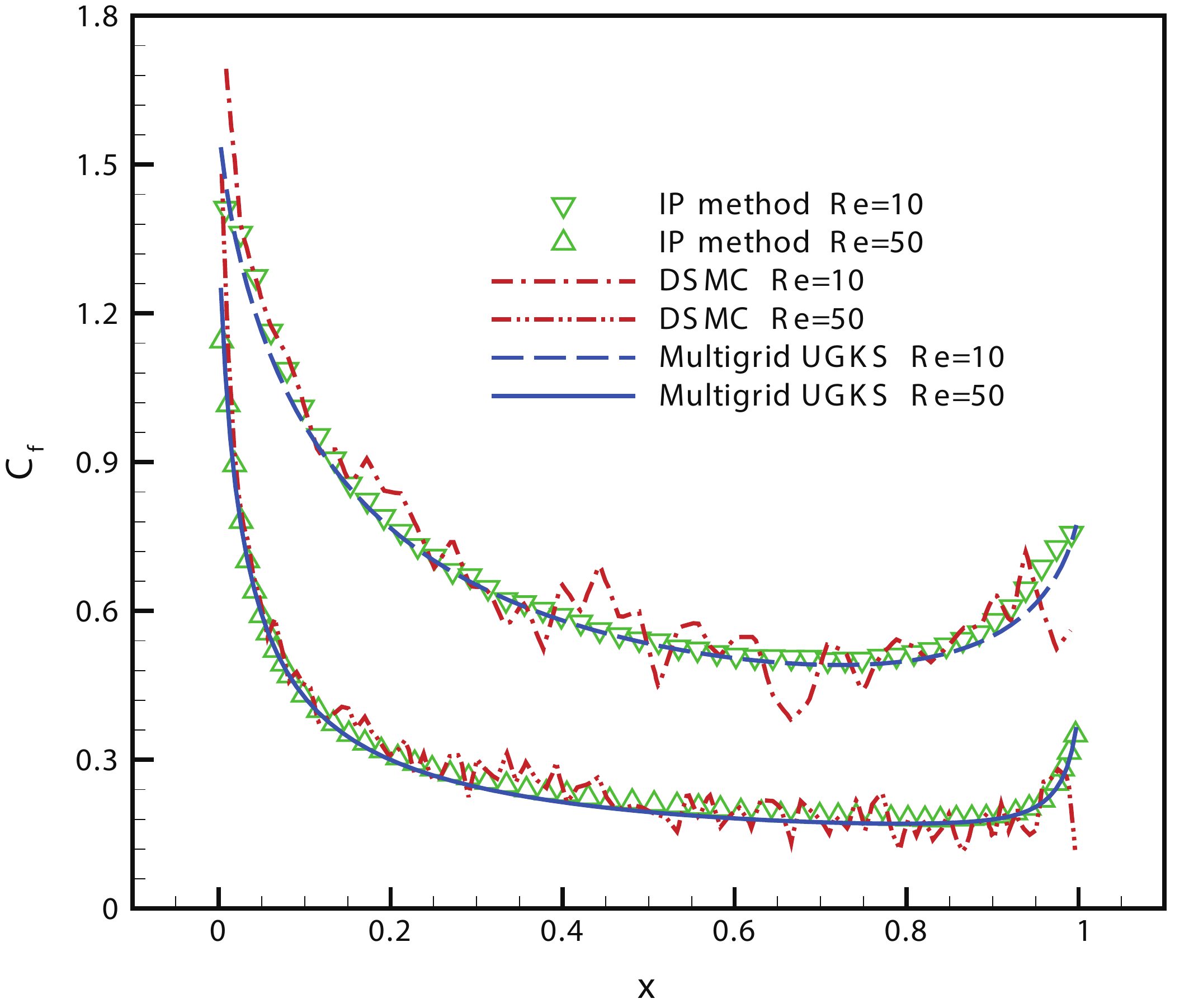}}
	\subfigure[\label{fig:flatPlateFrictionB}]{\includegraphics[width=0.48\textwidth]{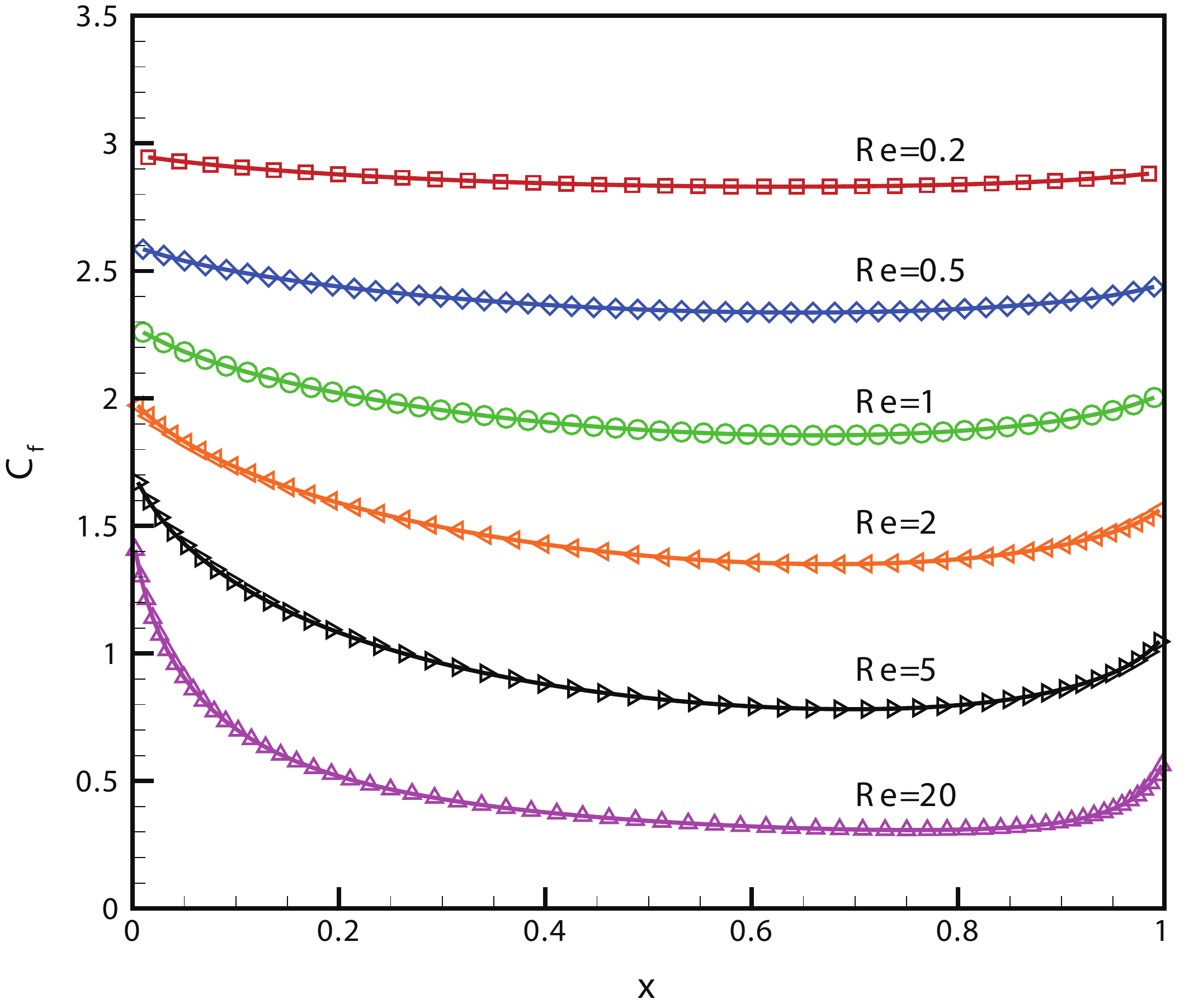}}
	\caption{\label{fig:flatPlateFriction}Distribution of the skin friction coefficients on the flat plate for the cases at different Reynolds numbers. (a) Re=10 and 50; symbols: IP results, red lines: DSMC data, blue lines: present results.  (b) the rest of numerical results; solid line: the multigrid UGKS, symbols: the implicit UGKS.}
\end{figure}

The distributions of the temperature around the plate for all cases are given in Fig.~(\ref{fig:flatPlateTemperature}). 
The current multigrid method can get consistent solutions with the implicit UGKS. It can be seen that the temperature varies more evidently in the cases at the lower Reynolds numbers than that at the higher Reynolds numbers. In order to compare with results of the DSMC method and the information preservation (IP) method obtained from paper\cite{sun2004drag}, the drag coefficient on the plate has been calculated by an integration of the skin friction coefficient over both sides of the plate, i.e.,
\begin{equation}
C_d = \dfrac{1}{L}\int_0^L{\left(C_f^{upper}+C_f^{lower}\right)}dl,
\end{equation}
where the skin coefficient is computed by
\begin{equation}
C_f = \dfrac{\tau_w}{\frac{1}{2}\rho_{\infty} U_{\infty}^2},
\end{equation}
in which the shear stress $\tau_w$ is the rate of the momentum transferred from gas to the solid wall for unit area. The drag coefficients varying with the Reynolds numbers are shown in Fig.~\ref{fig:flatPlateDrag}. 
The multigrid UGKS gives acceptable results matching well with the data from both the IP method and the DSMC. Specifically, the UGKS solutions match better with DSMC results in the cases with lower Reynolds numbers while get closer to IP method data in near continuum flow. The fitting formula as a reference in Fig.~\ref{fig:flatPlateDrag} is computed by
\begin{equation}
\ln (C_d \cdot Ma) = 0.225 - 0.333 \times {\ln}^2\left(\sqrt{Re}/{Ma}^{0.8}\right) + 0.031 \times {\ln}^3 \left(\sqrt{Re}/{Ma}^{0.8}\right).
\end{equation}
The skin friction coefficients distributed on the flat plate are shown in Fig.~\ref{fig:flatPlateFriction}, 
where the solutions at $Re=10$ and $50$ are compared with that from the DSMC and results in Fig.~\ref{fig:flatPlateFrictionA}, and the rest results are compared with those from the implicit UGKS in Fig.~\ref{fig:flatPlateFrictionB}. Good agreements have been obtained between the present results and the reference data.

\begin{figure}
	\centering
	\subfigure[]{\includegraphics[width=0.48\textwidth]{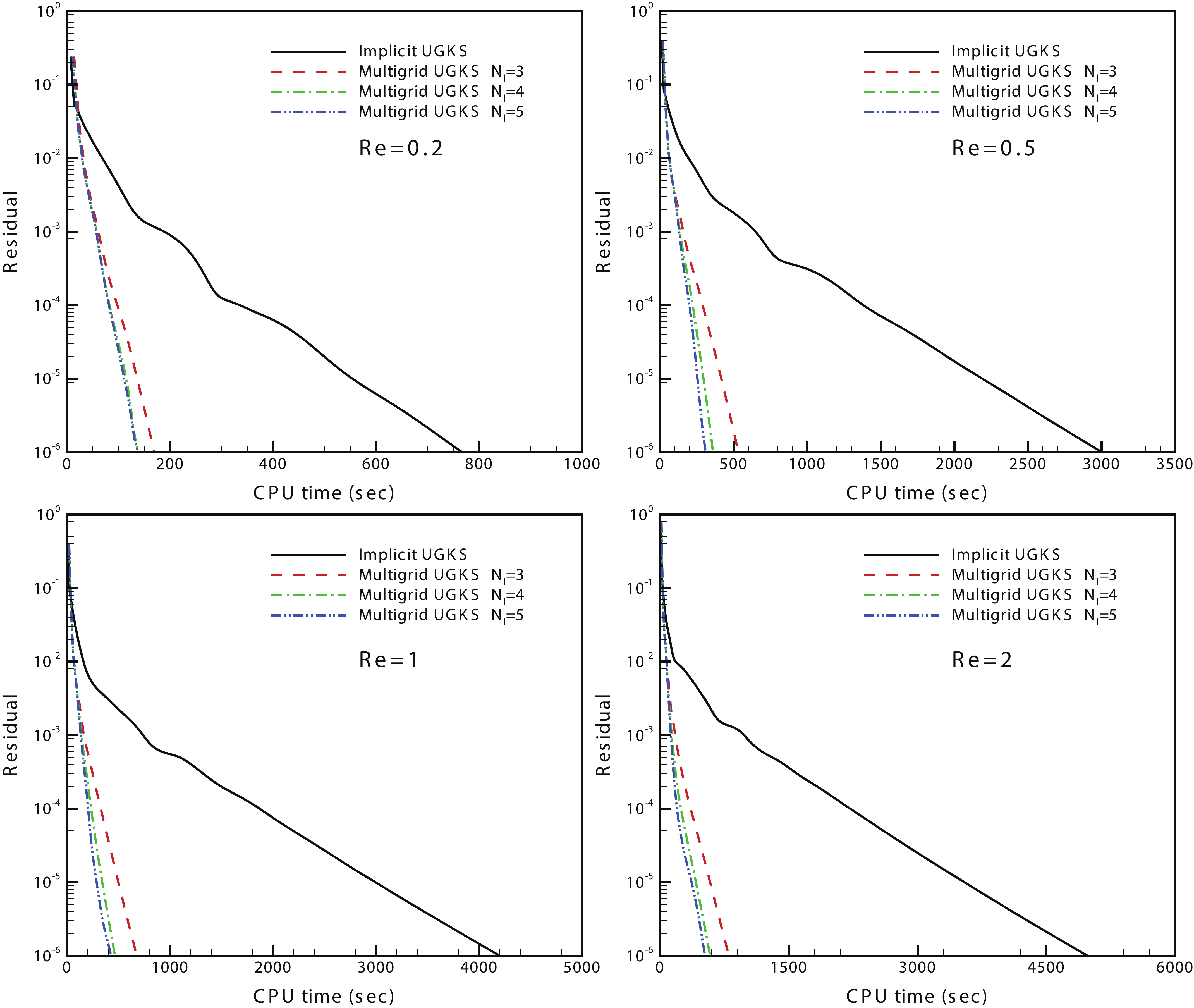}}
	\subfigure[]{\includegraphics[width=0.48\textwidth]{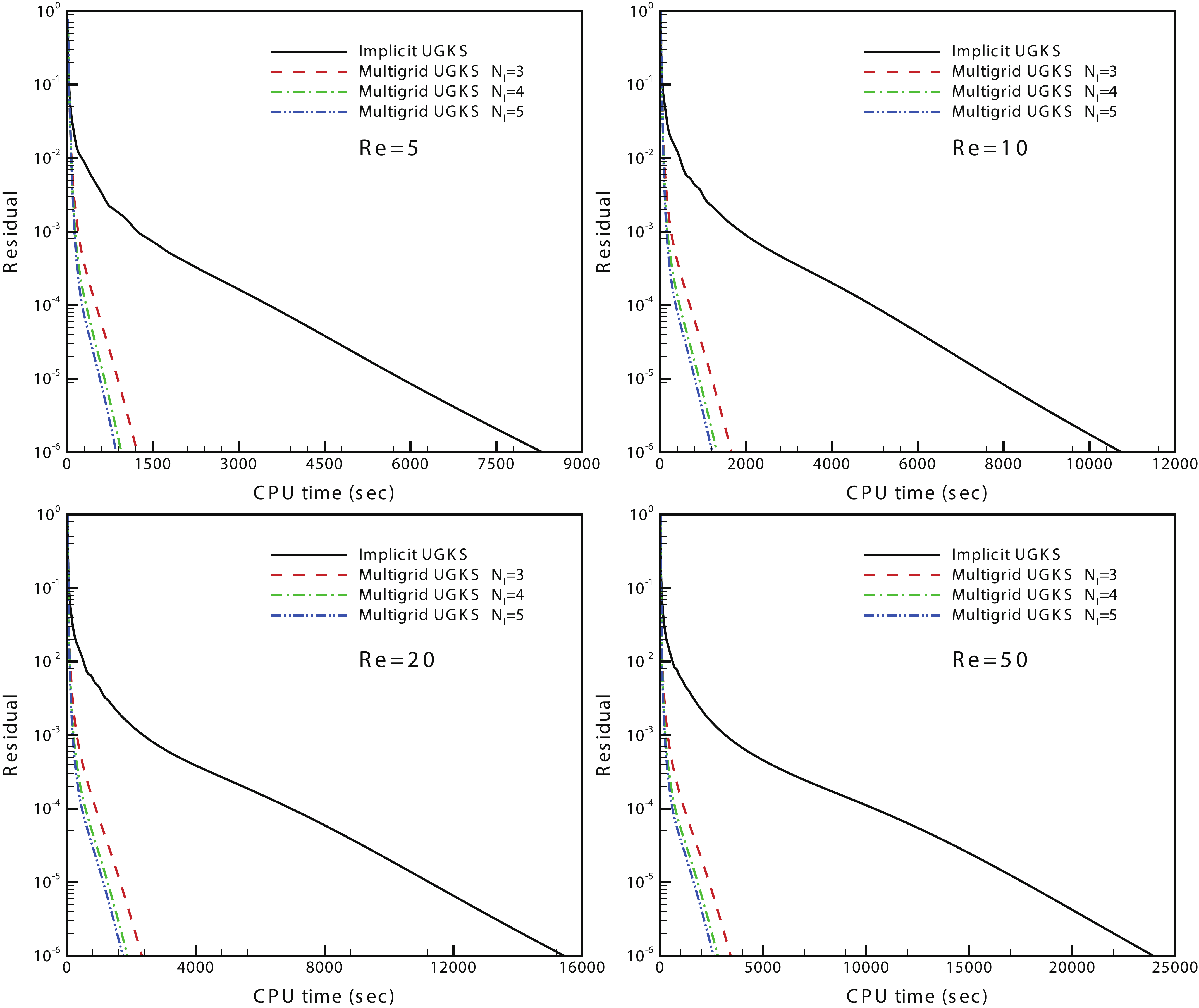}}
	\caption{\label{fig:flatPlateHistory}The convergence history of the implicit UGKS and the multigrid UGKS for the flow around the flat plate, indicated by the variation of the residual of X-momentum at different Reynolds numbers.}
\end{figure}

In order to illustrate the efficiency of the multigrid method,  we plot the convergence history of the implicit scheme and the multigrid method in Fig.~(\ref{fig:flatPlateHistory}).
The calculation stops when the mean squared residuals as defined in Eq.~(\ref{eq:Residual_Def}) get lower than $1.0 \times 10^{-6}$. For a better description, the iteration steps and the total CPU time that cost by the implicit UGKS, three-level, four-level and five-level multigrid UGKS are listed in Table~\ref{tab:flatPlateIteration}
\begin{table}
	\caption{\label{tab:flatPlateIteration}Total iteration steps cost by the implicit UGKS and multigrid methods.}
	\begin{ruledtabular}
		\begin{tabular}{cccccc}
			\multirow{2}{*}{Re}	 & \multirow{2}{*}{IUGKS} 
			& \multicolumn{3}{c}{MIUGKS} & \multirow{2}{*}{Speedup} \\ 
			\cline{3-5}
			~ & ~ & $N_l$=3 & $N_l=4$ & $N_l=5$ & ~\\
			\hline
			0.2 &  118 &  18 &  14 &  14 &  8.4\\
			0.5 &  245 &  30 &  20 &  18 & 13.6\\
			1   &  334 &  36 &  24 &  23 & 14.5\\
			2   &  504 &  52 &  37 &  34 & 14.8\\
			5   &  841 &  79 &  61 &  55 & 15.3\\
			10  & 1285 & 127 & 100 &  92 & 14.0\\
			20  & 1845 & 176 & 142 & 130 & 14.2\\
			50  & 2854 & 260 & 211 & 195 & 14.6\\
		\end{tabular}
	\end{ruledtabular}
\end{table}
and Table~\ref{tab:flatPlateCPUtime}.
\begin{table}
\caption{\label{tab:flatPlateCPUtime}Total CPU time (min) cost by the implicit UGKS and multigrid methods.}
\begin{ruledtabular}
\begin{tabular}{cccccc}
\multirow{2}{*}{Re}	& \multirow{2}{*}{IUGKS} 
& \multicolumn{3}{c}{MIUGKS} & \multirow{2}{*}{Speedup} \\ 
\cline{3-5}
~ & ~ & $N_l$=3 & $N_l=4$ & $N_l=5$ & ~\\
\hline
0.2 &  13.5 &  3.3 &  2.5 & 2.5 & 5.3 \\
0.5 &  51.5 &  9.8 &  6.6 & 6.0 & 8.6 \\
1   &  70.2 & 11.8 &  8.0 & 7.6 & 9.2\\
2   &  83.0 & 13.5 &  9.7 & 8.9 & 9.3\\
5   & 138.6 & 20.5 & 16.0 &14.4 & 9.6\\
10  & 179.2 & 27.9 & 22.1 &20.4 & 8.8\\
20  & 257.2 & 38.7 & 31.5 &28.9 & 8.9\\
50  & 398.8 & 57.1 & 46.7 &43.2 & 9.2\\
\end{tabular}
\end{ruledtabular}
\end{table}
Generally, the multigrid method can speed up the computation by about $9$ times for the case with five levels of grids. From these tables it can be known that the CPU time cost by the multigrid method (e.g., with 5-level grids) is about 57\% more than the implicit UGKS, due to the consumption on the coarser grids and interpolation between different grid levels, and the multi-smoothing process on each level. In the current test case, the convergence speed gets reduced with the increase of the Reynolds number.

\subsection{Hypersonic flow past a square cylinder}\label{sec:square}
Following the research in paper \cite{chen2017reduction}, the multigrid implicit UGKS is used to study hypersonic flow 
around a square cylinder. The freestream is argon gas at $Ma=5$ with an initial temperature of $T_{\infty}=273K$. The Knudsen number of the freestream is $0.1$, defined relative to the diameter of the square cylinder by the VHS model with $\omega=0.81$. The solid surface of the square cylinder is taken as isothermal wall with a fixed temperature of $T_w = 273K$. Due to the symmetric property, only half of the physical domain is considered.

\begin{figure}
	\centering
	\subfigure[\label{fig:squareMesh1}]{\includegraphics[width=0.32\textwidth]{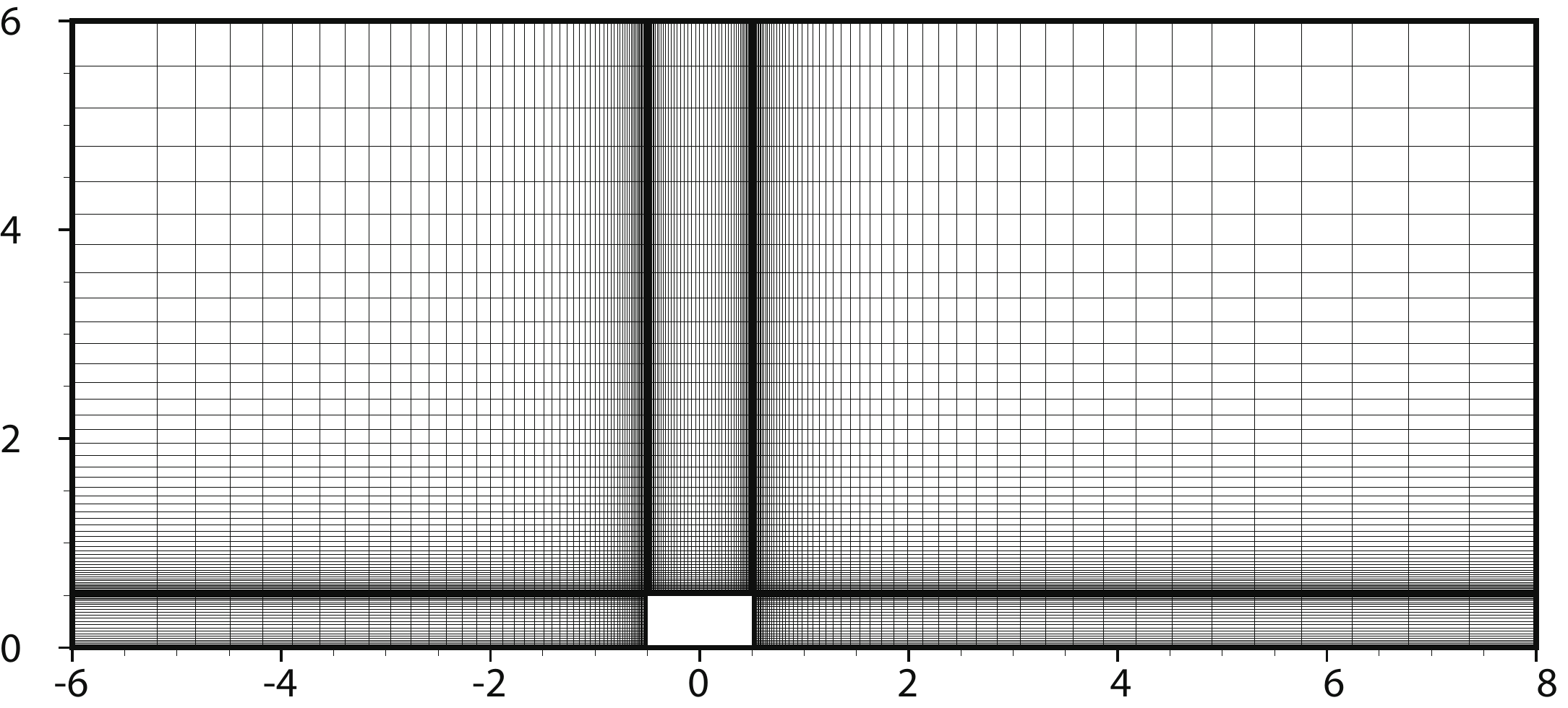}}
	\subfigure[\label{fig:squareMesh2}]{\includegraphics[width=0.32\textwidth]{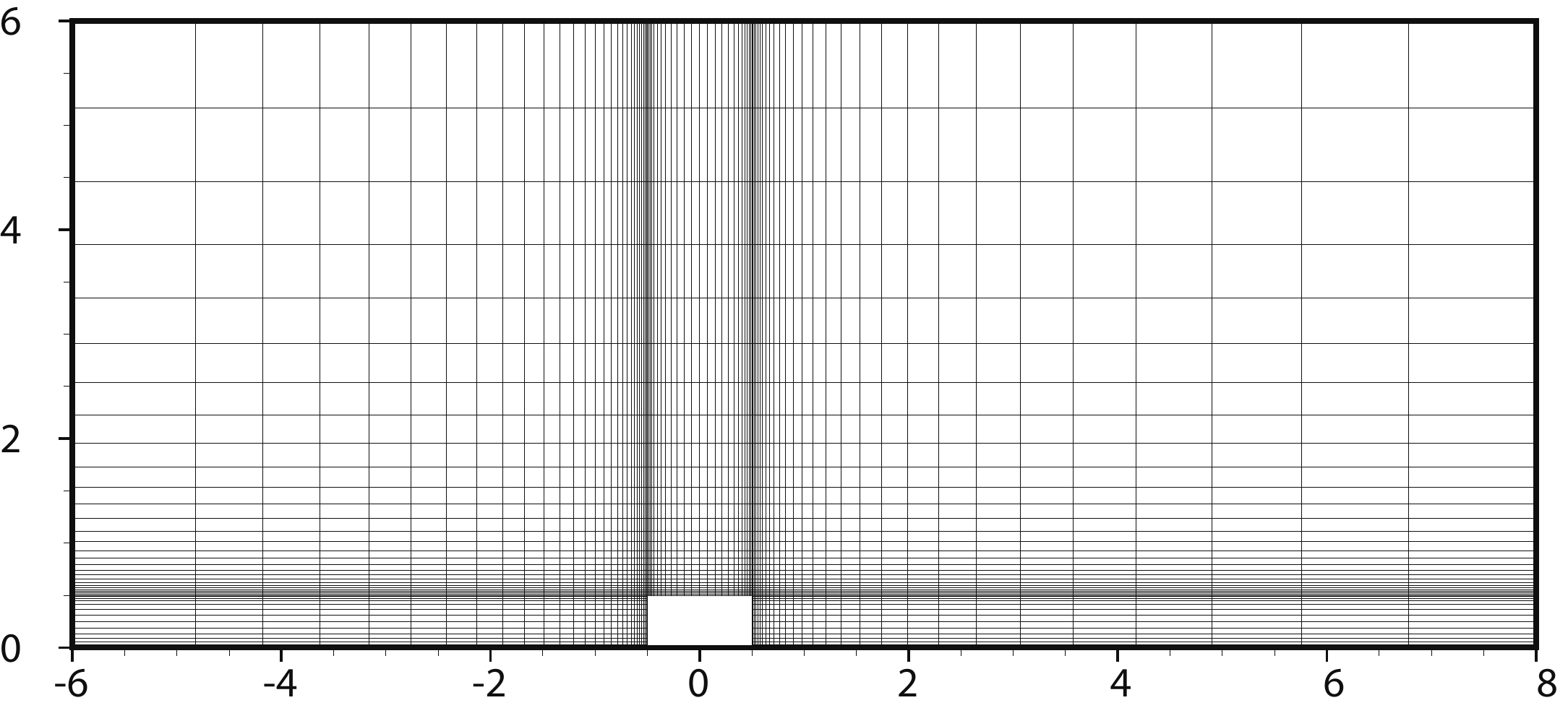}}
	\subfigure[\label{fig:squareMesh3}]{\includegraphics[width=0.32\textwidth]{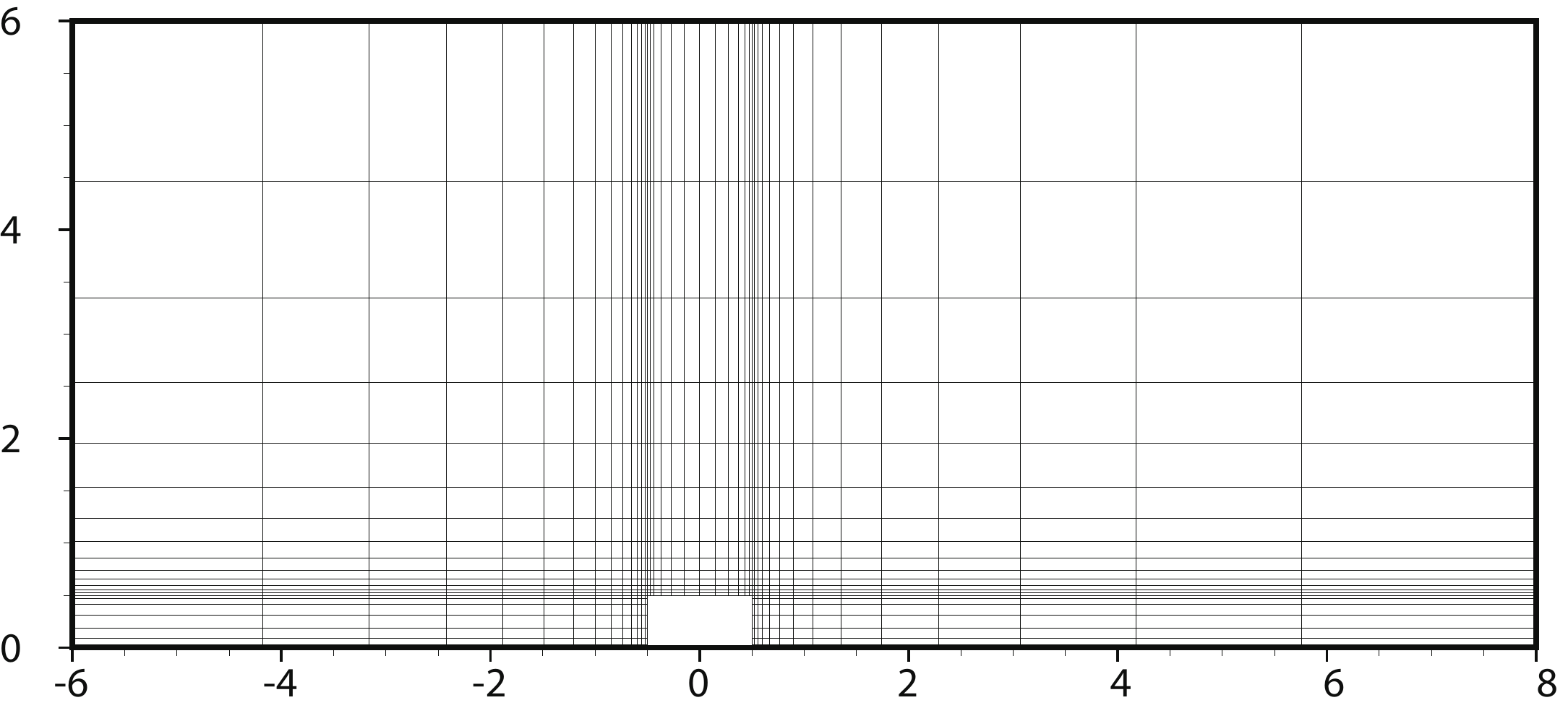}}
	\caption{\label{fig:squareMesh}Multiple grids used in the case of hypersonic flow around the square cylinder.}
\end{figure}

The physical domain is discretized into $12096$ cells with a nearest distance of $0.005m$ to the square surface. The multiple grids used in this case are shown in Fig.~\ref{fig:squareMesh}. 
In velocity space, $101 \times 101$ discrete velocity points are used for the integration of distribution function by the Newton-Cotes rules. Since a sudden start of a  hypersonic flow in the whole computation domain with the same speed imposes great challenges in the initial simulation at the rear part of the square cylinder, in this case we initialize the rear domain behind the cylinder with $\rho_r = 0.1\rho_\infty$ and $U_r=0$ initially for the convergence evolution.

\begin{figure*}
	\centering
	\subfigure[\label{fig:squareTemperature}]{\includegraphics[width=0.32\textwidth]{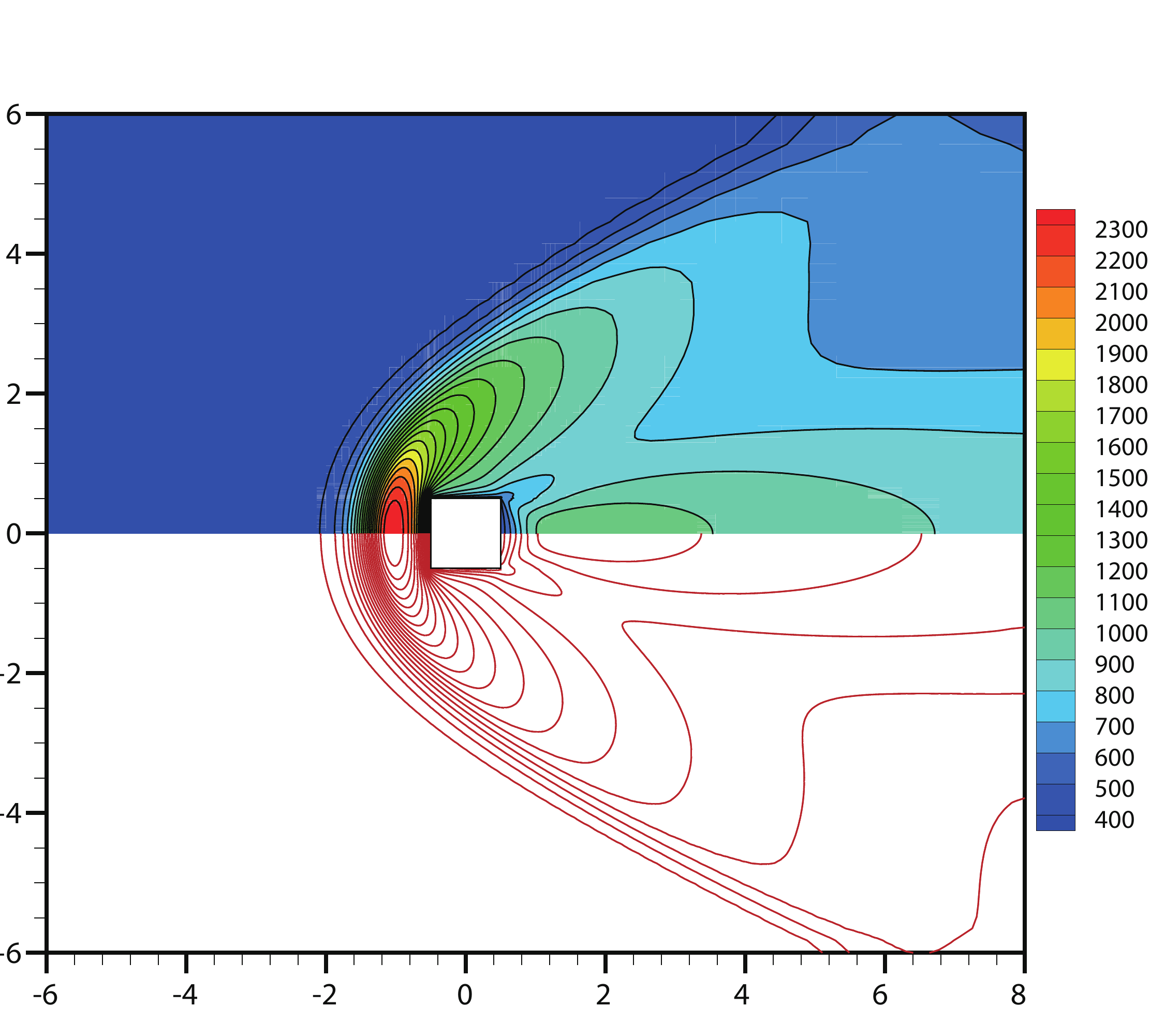}}
	\subfigure[\label{fig:squareVelocityX}]{\includegraphics[width=0.32\textwidth]{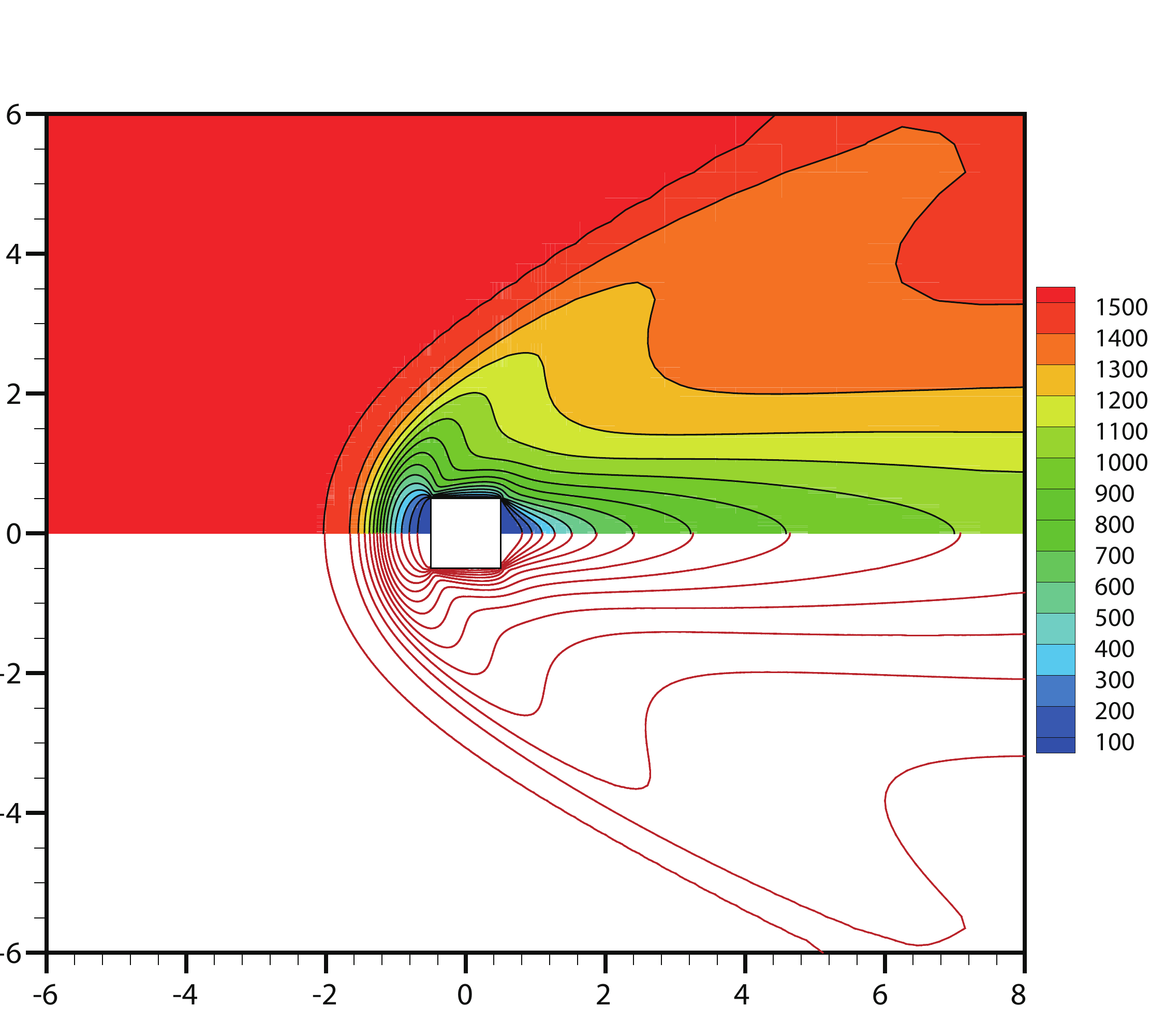}}
	\subfigure[\label{fig:squareVelocityY}]{\includegraphics[width=0.32\textwidth]{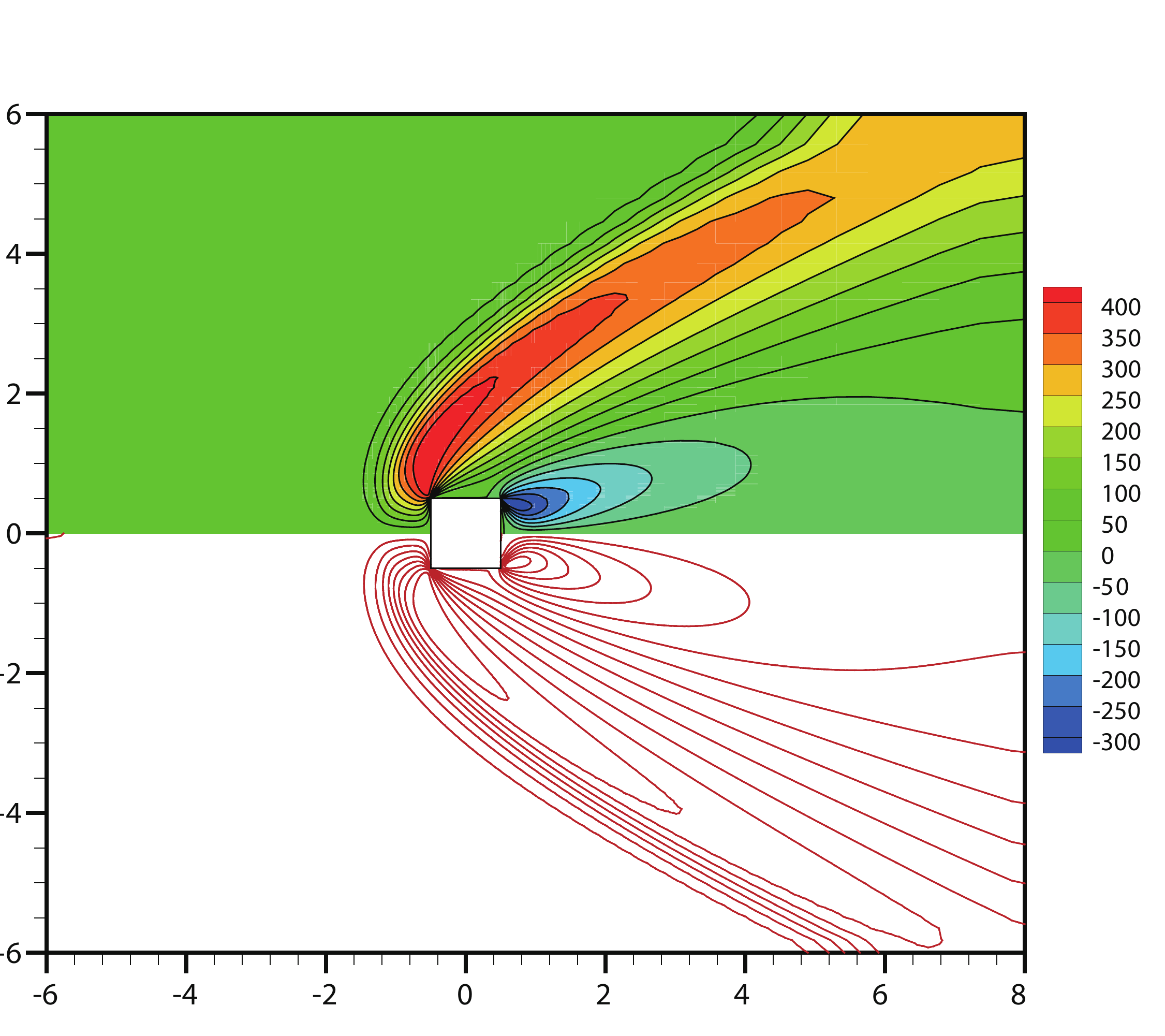}}
	\caption{\label{fig:squareFlowField}Steady state of the hypersonic flow around the square cylinder at $Ma=5$ for (a) temperature, (b) X-velocity and (c) Y-velocity. Upper: the current multigrid UGKS. Lower: DSMC data computed by dsmcFoam \cite{scanlon2010open}.}
\end{figure*}

\begin{figure*}
	\centering
	\subfigure[\label{fig:squareWallPressure}]{\includegraphics[width=0.32\textwidth]{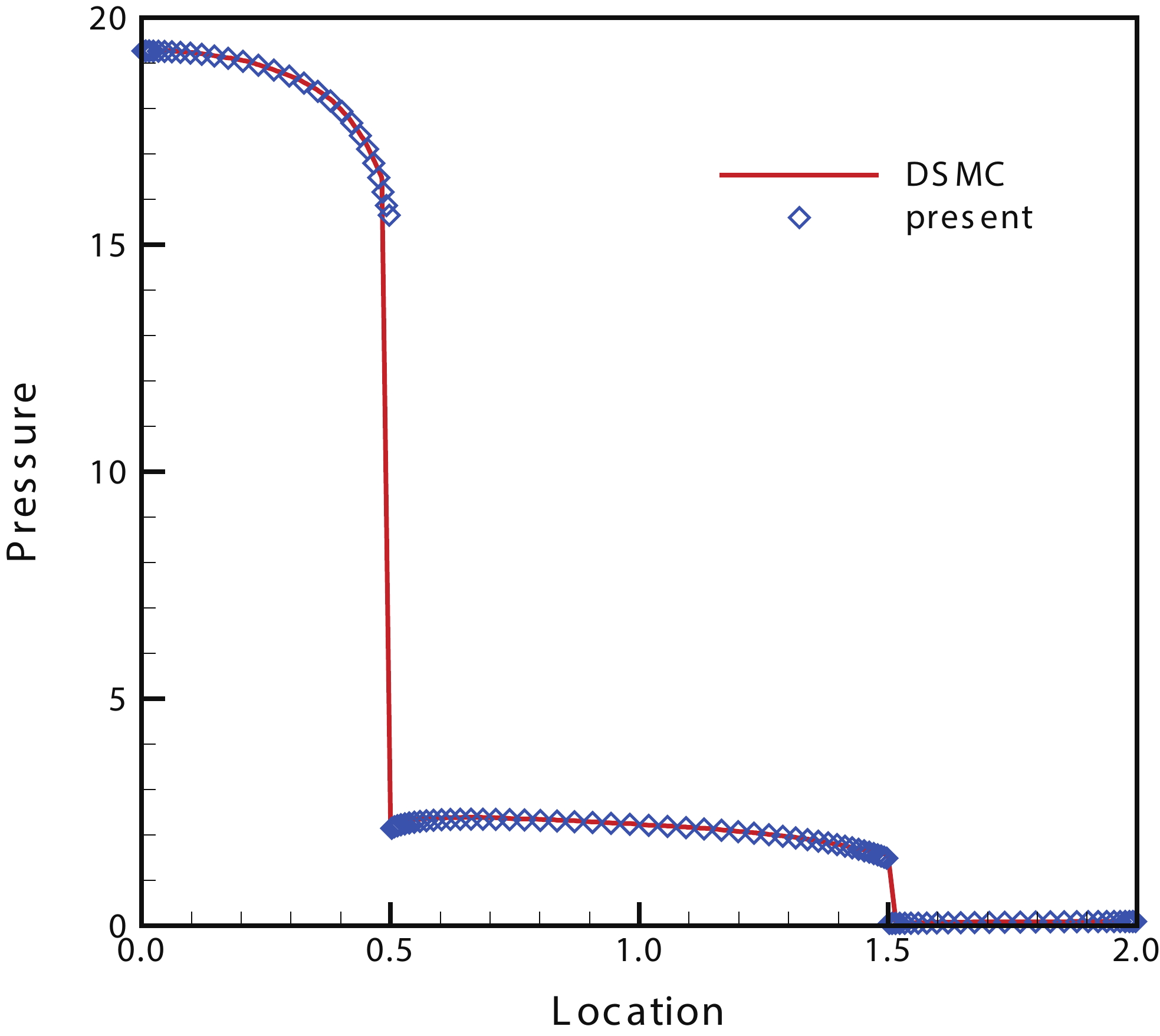}}
	\subfigure[\label{fig:squareWallStress}]{\includegraphics[width=0.32\textwidth]{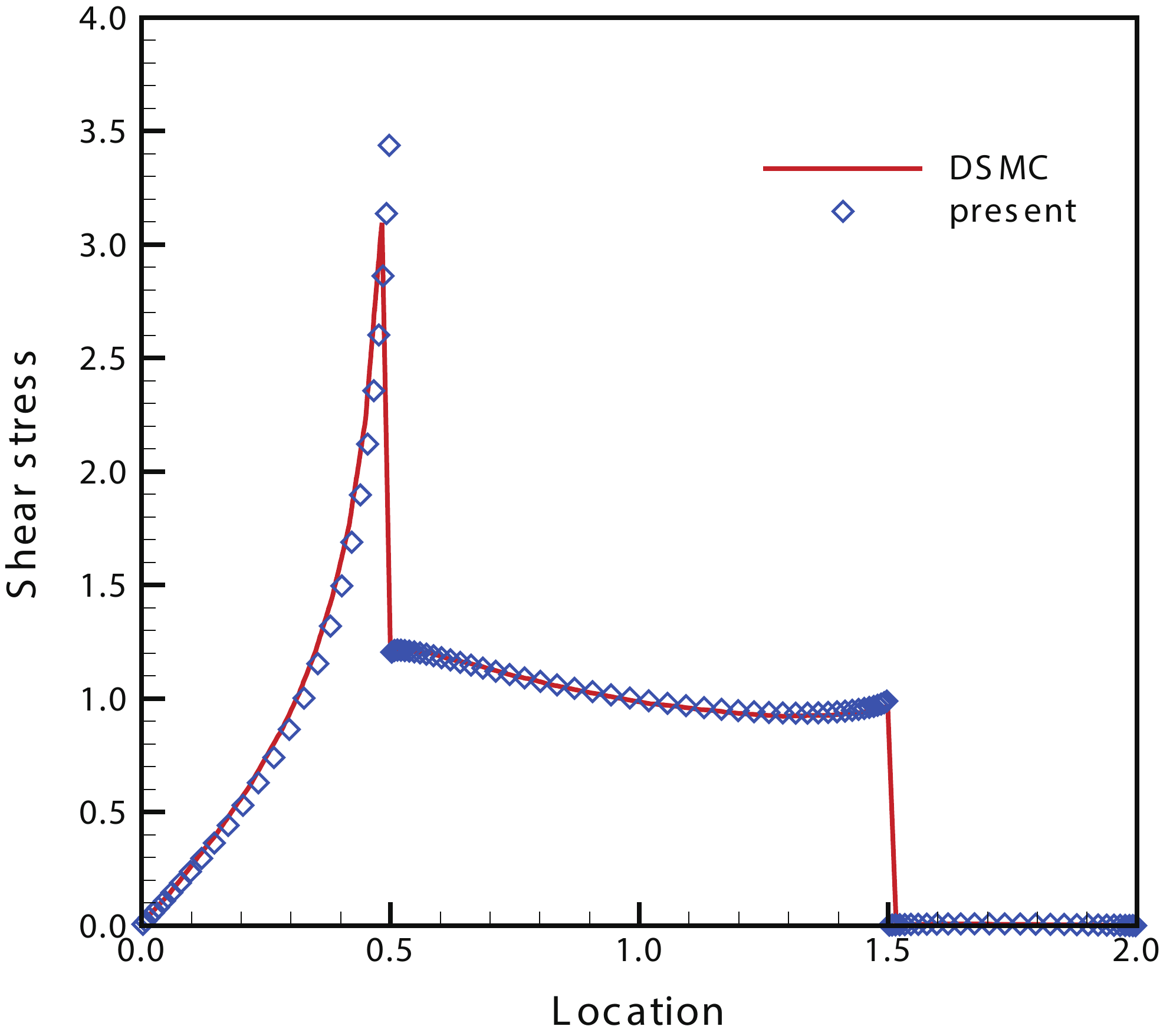}}
	\subfigure[\label{fig:squareWallHeatflux}]{\includegraphics[width=0.32\textwidth]{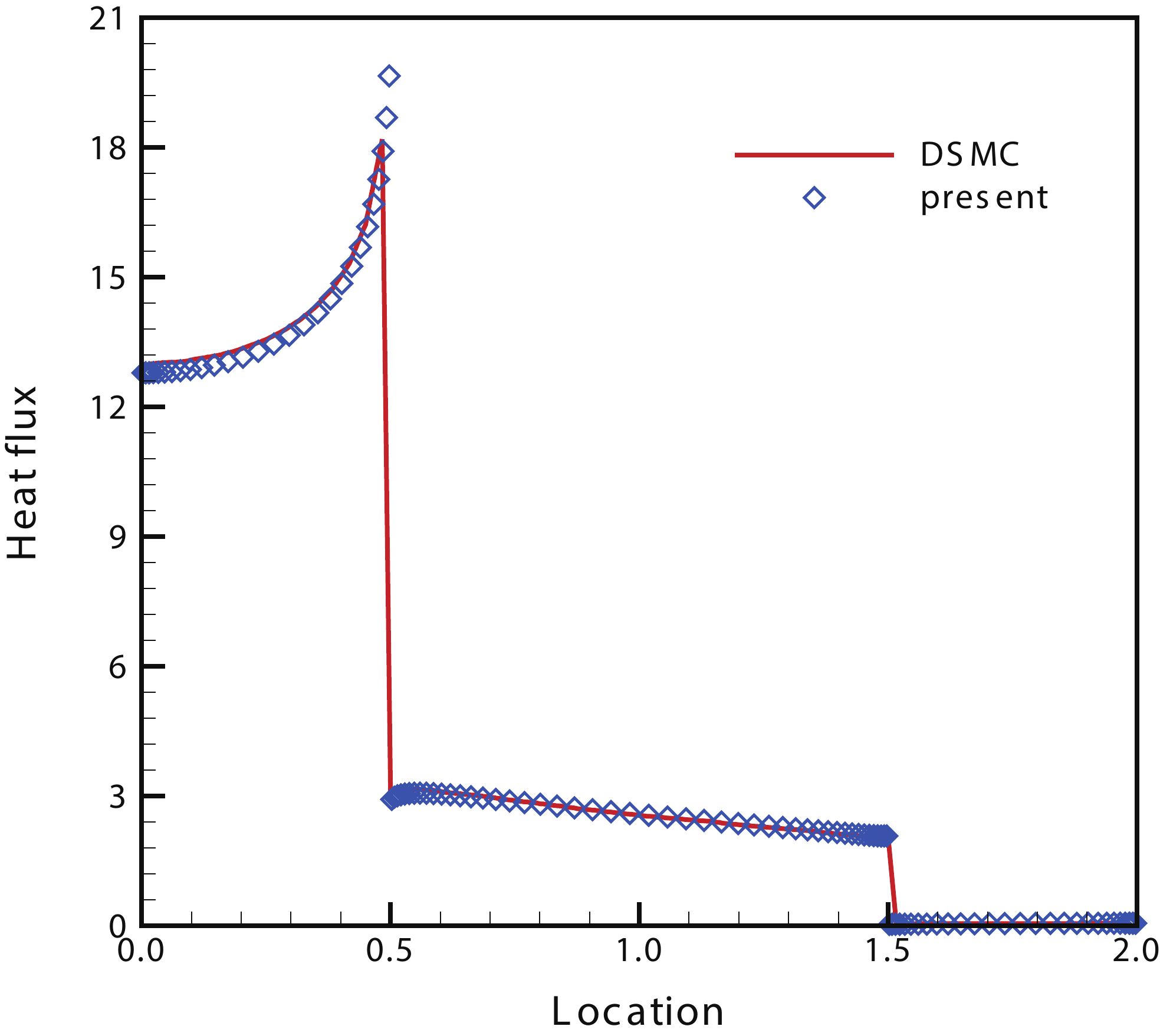}}
	\caption{\label{fig:squareWall}Distribution of the surface quantities along the square cylinder wall from the stagnation point to the trailing edge, (a) pressure, (b) shear stress and (c) heat flux. The pressure and shear stress are normalized by $\rho_{\infty} C_{\infty}^2$, and the heat flux is normalized by $\rho_{\infty} C_{\infty}^3$ where $C_{\infty} = \sqrt{2 R T_{\infty}}$. }
\end{figure*}

\begin{figure*}
	\centering
	\subfigure[\label{fig:squareUpstreamDensity}]{\includegraphics[width=0.32\textwidth]{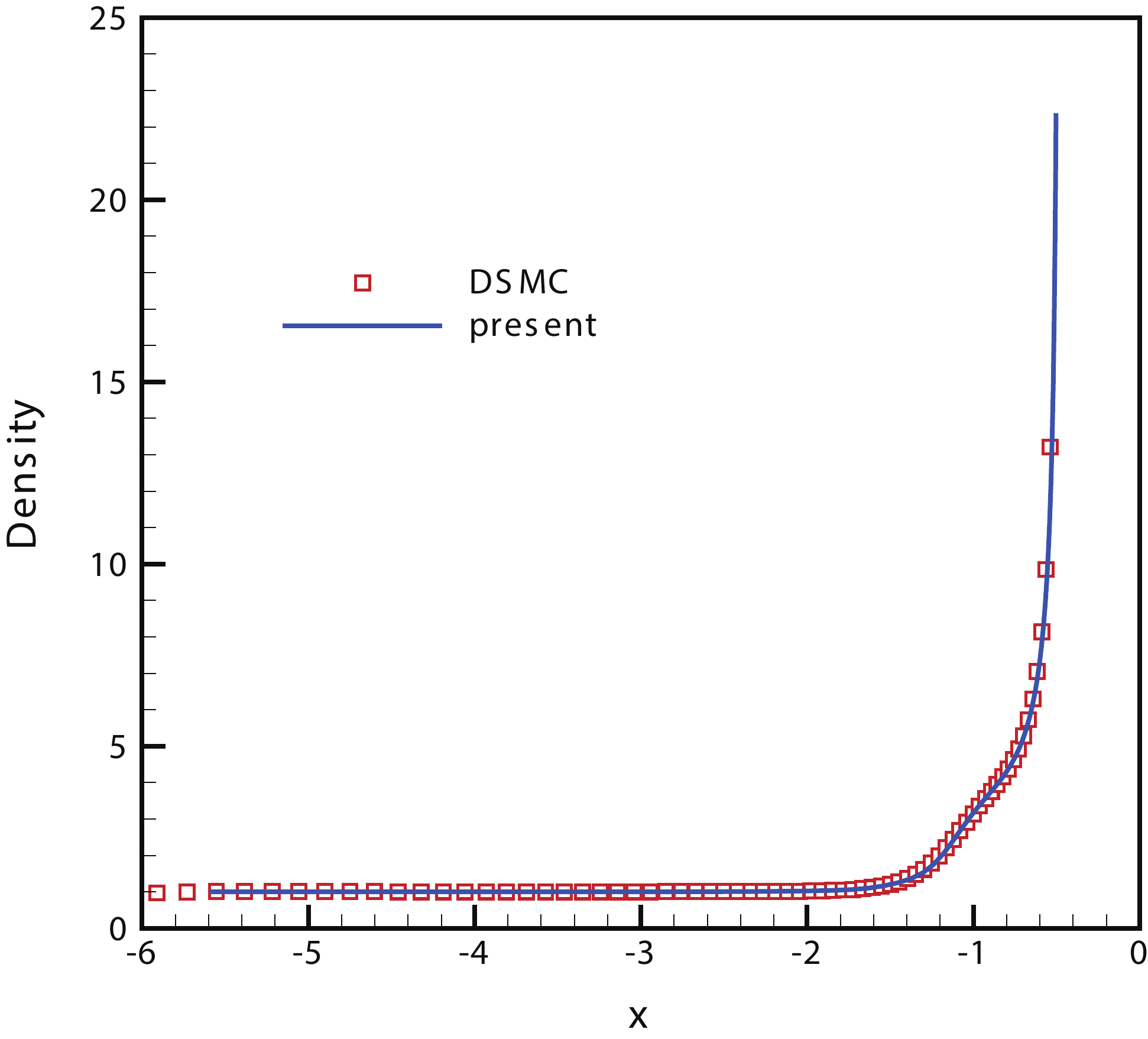}}
	\subfigure[\label{fig:squareUpstreamVelocity}]{\includegraphics[width=0.32\textwidth]{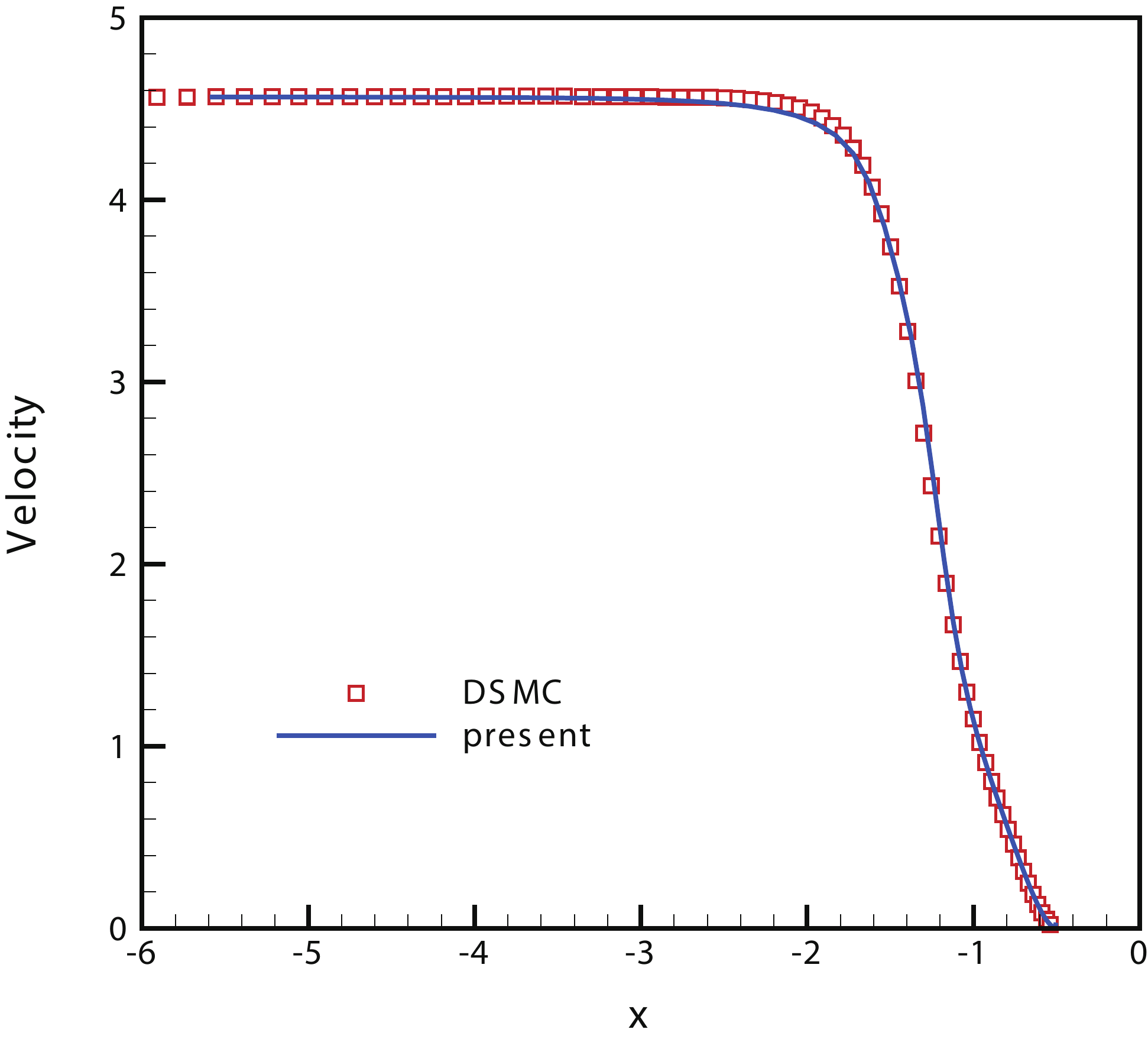}}
	\subfigure[\label{fig:squareUpstreamTemperature}]{\includegraphics[width=0.32\textwidth]{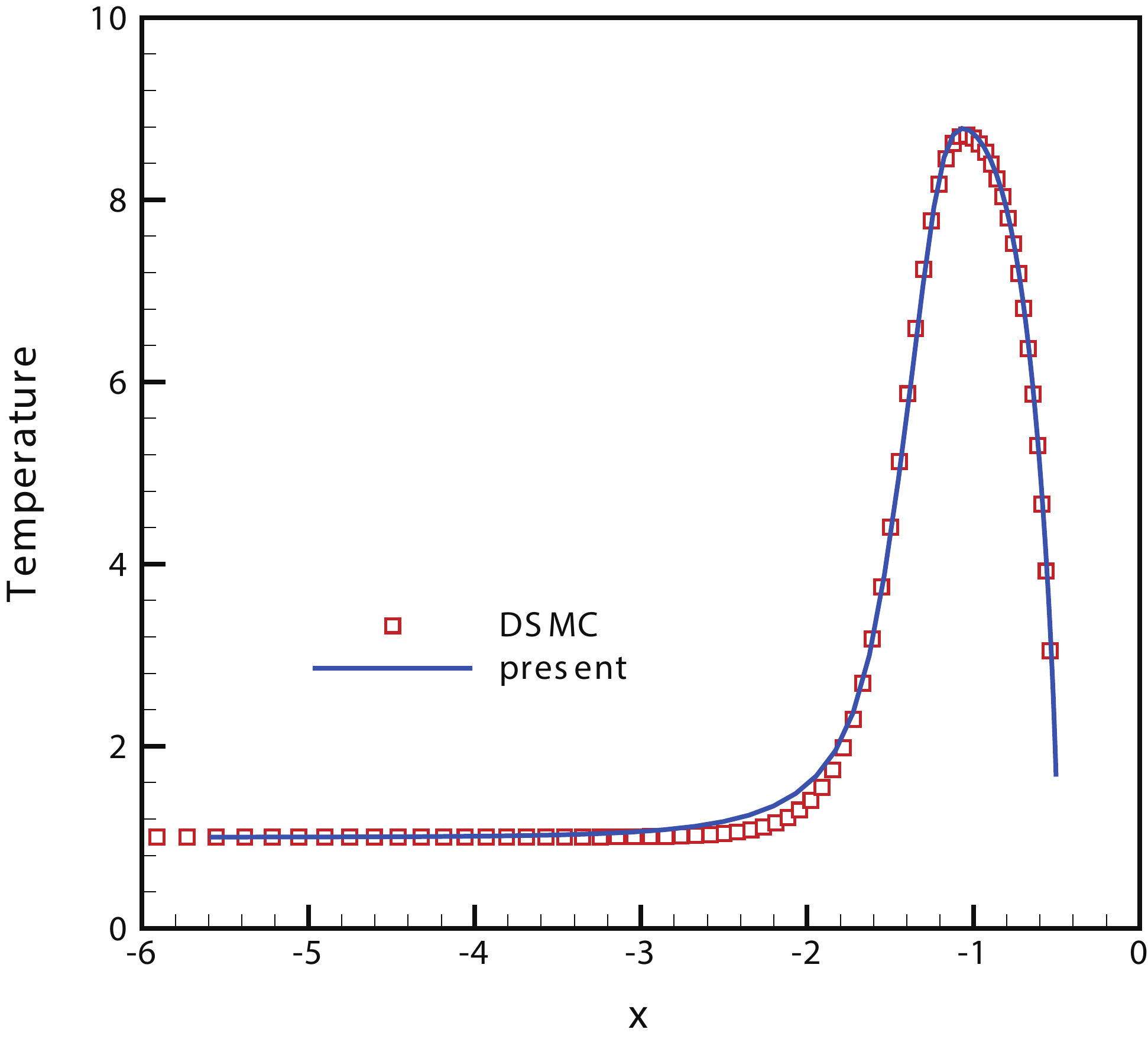}}
	\caption{\label{fig:squareUpstream}Flow variables along the symmetric axis in the upstream for (a) density, (b) velocity and (c) temperature. The density, velocity and temperature are normalized by $\rho_{\infty}$, $\sqrt{2 R T_\infty}$ and $T_{\infty}$ respectively.}
\end{figure*}

Fig.~\ref{fig:squareFlowField}
shows the flow field at the steady state, including the distributions of temperature, horizontal velocity and vertical velocity. The results of the multigrid UGKS are compared with the DSMC results obtained by dsmcFoam in \cite{chen2017reduction}. The present results agree well with the reference ones for each contour. The normalized surface quantities, such as normal pressure, shear stress, and heat flux, are plotted in Fig.~\ref{fig:squareWall}.
And the distribution of the flow variables along the symmetric axis in the upstream are presented in Fig.~\ref{fig:squareUpstream} 
and compared with DSMC results. Basically, the present results obtained from the multigrid UGKS match well with the reference data.

\begin{figure}
	\centering
	\includegraphics[width=0.6\textwidth]{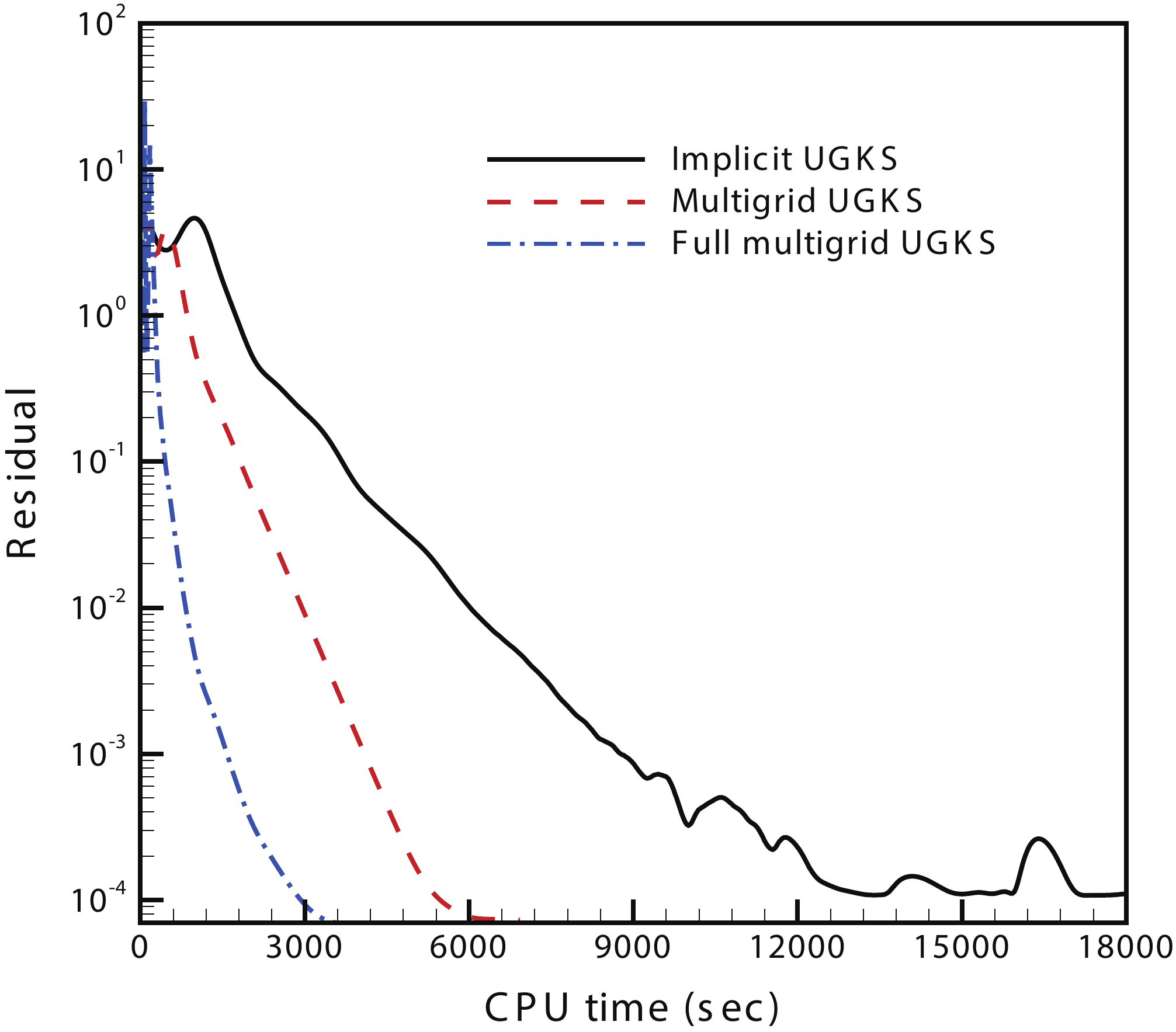}
	\caption{\label{fig:squareHistory} Convergence history of the flow around a square cylinder at $Ma=5$.}
\end{figure}

For this case, the numerical time step is employed, which increases exponentially with iteration steps by
\begin{displaymath}
\Delta t_n = a^n \Delta t_p.
\end{displaymath}
During calculations, it is found that the multigrid method with multiple smoothing steps is more robust than the original implicit UGKS. So $a=3$ and $a=1.2$ are used for the multigrid UGKS and the implicit scheme, respectively. The convergence history is plotted in Fig.~\ref{fig:squareHistory}.
In this case, the full multigrid method is used to give a better initial approximate solution for finer grids. It can be observed that the full multigrid method does improve the convergence efficiency and it is about $5$ times faster than the implicit UGKS on a single grid for this case. For the same case, the DSMC method obtaioned by dsmcFoam \cite{scanlon2010open} takes about $8$ hours to get the steady state solution by parallel computing on two server nodes, each of which has two processors of Intel(R) Xeon(R) E5-2680 v3@2.5GHz with 12 cores. However, the multigrid implicit UGKS needs only $50$ minutes by serially computing on a single machine with a CPU of Intel(R) Core(TM) i5-4570 CPU@3.2GHz. The improvement of the efficiency in the current scheme in comparison with the DSMC method is significant. In the current high speed flow study, the MIUGKS is about $300$ times faster than the DSMC.

\section{Conclusions}\label{sec:conclusions}
In this paper, we present a geometric multigrid method for the implicit unified gas-kinetic scheme for rarefied flow computations. Both stages in the implicit UGKS, i.e., the prediction of equilibrium state and the evolution of distribution function, are treated with multigrid techniques. In prediction step, the governing equations of macroscopic conservative flow variables are solved by the full approximation scheme on each level of grid. While in the evolution step, the governing equations of microscopic distribution function are solved by the correction scheme, by which the distribution function is not required on coarser grids so that the increasing of the computational cost can be well controlled. With a recursive definition of the FAS cycle and CS cycle, the multigrid method for the implicit UGKS has been constructed from the two-grid method.

The multigrid implicit UGKS has been applied to the study of non-equilibrium flows, such as lid-driven cavity flow at different Knudsen numbers, subsonic flow around a plate, and the hypersonic flow past a square cylinder, and the accuracy and efficiency of the multigrid method has been well demonstrated. In comparison with the implicit UGKS with a single level of grid, which is already hundreds times faster than the explicit UGKS, the convergence efficiency of multigrid implicit UGKS has been further improved in all flow regimes from low and high speed flows. In general, the multigrid UGKS with $5$-level grids takes about $57$\% more CPU time than the implicit UGKS with a single level of grid within one iteration step, but overall it is about $5$ to $9$ times more efficient than the implicit scheme. As a further development, the AMG technique can be employed here as well to remove the generation of multiple grids. The multigrid UGKS can be also developed on unstructured mesh for the applications with complex geometry.

For rarefied flow computations, especially for the hypersonic flow, the DSMC is currently the dominant method
in the engineering rarefied flow applications. However, with the implementation of the implicit and multigrid techniques, the UGKS becomes more efficient than the DSMC method, at least in all cases presented in this paper. Even for the high speed flow at Mach number $5$ and Knudsen number $0.1$, the UGKS is two orders of magnitude more efficient than the DSMC method. For low speed flows in the transition and near continuum regimes, the efficiency differences between UGKS and DSMC get even larger. With the implementation of acceleration techniques, such as implicit, preconditioning, local time, and multigrid, the UGKS becomes an accurate, reliable, and efficient method  for rarefied flow computations. It has been successfully used in rarefied flow applications \cite{jiangdingwu}, and been extended to other non-equilibrium transport processes, such as radiative transfer and plasma \cite{sun2017radiative,liu2016plasma}. With computational efficiency increase, the equation-based flow solver will become an alternative choice in the non-equilibrium flow study and practical engineering applications.

\begin{acknowledgments}
The authors would like to thank Mr. Lianhua Zhu for providing the DSMC results for hypersonic flow around the square cylinder,
and the detailed computational cost.
This work of Zhu and Zhong was supported by National Natural Science Foundation of China (Grant No. 11472219) and National Pre-Research Foundation of China, as well as the 111 Project of China (B17037).
The research work of Xu is supported by Hong Kong research grant council (16207715,16211014,620813) and NSFC (91330203, 91530319).
\end{acknowledgments}

\bibliography{mybibfile}

\end{document}